\documentclass[12pt]{JHEP3} 
\usepackage{graphicx,caption}
\usepackage{amsmath,amssymb,amsfonts}
\usepackage[T1]{fontenc} 
\usepackage[latin1]{inputenc}
\usepackage{bbm,mathrsfs}
\usepackage{cite}

\newlength{\textlength}
\newlength{\overlinelength}
\newcommand{\ovl}[2][.55]{\settowidth{\textlength}{$#2$}
  \setlength{\overlinelength}{0.1pt}
  \addtolength{\overlinelength}{0.75\textlength}
  \makebox[\textlength][s]{$#2$} \hspace{-.55\textlength}
  \hspace{-\overlinelength}\hspace{#1\overlinelength}
  \overline{\makebox[\overlinelength][s]{\vphantom{$#2$}}}
  \hspace{-#1\overlinelength}\hspace{.55\textlength}}

\def\lsim{\mathrel{\rlap{\lower4pt\hbox{\hskip1pt$\sim$}}
    \raise1pt\hbox{$<$}}}                

\newcommand{\abs}[1]{\left| #1\right|} 
\newcommand{\bm}{\ovl{\mu}}            
\newcommand{\rb}{\ovl{\rho}}           
\newcommand{\rt}{\widetilde{\rho}}     
\newcommand{\tm}{\widetilde{\mu}}      
\newcommand{\tM}{\widetilde{M}}        

\DeclareMathOperator{\diag}{diag}
\DeclareMathOperator{\im}{Im}

\DeclareMathOperator{\tr}{tr}

\title{CP Violation and Neutrino Masses and Mixings from Quark Mass
  Hierarchies}

\author{Wilfried Buchm\"uller
  \\
  Deutsches Elektronen-Synchrotron DESY, Hamburg, Germany}

\author{Laura Covi
  \\
  Deutsches Elektronen-Synchrotron DESY, Hamburg, Germany}

\author{David Emmanuel-Costa
  \\
  CFTP, Departamento de Fisica, Istituto Superior Tecnico, Lisbon,
  Portugal}

\author{S\"oren Wiesenfeldt
  \\
  Department of Physics, University of Illinois at Urbana-Champaign,
  USA}

\abstract{We study the connection between quark and lepton mass matrices
  in a supersymmetric SO(10) GUT model in six dimensions, compactified
  on an orbifold.  The physical quarks and leptons are mixtures of brane
  and bulk states.  This leads to a characteristic pattern of mass
  matrices and high-energy CP violating phases.  The hierarchy of up and
  down quark masses determines the CKM matrix and most charged lepton
  and neutrino masses and mixings.  The small hierarchy of neutrino
  masses is a consequence of the mismatch of the up and down quark mass
  hierarchies.  The effective CP violating phases in the quark sector,
  neutrino oscillations and leptogenesis are unrelated.  In the neutrino
  sector we can accomodate naturally $\sin \theta_{23} \sim 1$, $\sin
  \theta_{13} \lesssim 0.1$ and $m_1 \lesssim m_2 \sim \sqrt{\Delta
    m^2_\text{sol}} < m_3 \sim \sqrt{\Delta m^2_\text{atm}}$.}

\keywords{CP violation, Field Theories in Higher Dimensions, GUT,
  Neutrino Physics}

\preprint{DESY 07-141}

\begin{document}

\section{Introduction}\label{se:intro}

Grand unified theories (GUTs) appear to be the most promising framework
\cite{Fritzsch:1999ee,Ross:2000fn} to address the still challenging
question of quark and lepton masses and mixings.  During the past years
new results from neutrino physics have shed new light on this problem,
and the large differences between the mass hierarchies and mixing angles
of quarks, charged leptons and neutrinos impose strong constraints on
unified extensions of the Standard Model (SM)
\cite{Mohapatra:2006gs,Altarelli:2006ri}.  Massive neutrinos are most
easily incorporated in theories with right-handed neutrinos, which leads
to SO(10) as preferred GUT gauge group
\cite{Georgi:1975qb,Fritzsch:1974nn}.
 
Higher-dimensional theories offer new possibilities to describe gauge
symmetry breaking, the notorious doublet-triplet splitting and also
fermion masses.  A simple and elegant scheme is provided by orbifold
compactifications which have recently been considered for GUT models in
five and six dimensions
\cite{Kawamura:2000ev,Altarelli:2001qj,Hall:2001pg,Hebecker:2001wq,Asaka:2001eh,Hall:2001xr}.
In this paper we analyse in detail the connection between quark and
lepton mass matrices in the six-dimensional (6D) GUT model suggested in
\cite{Asaka:2003iy}, for which also proton decay
\cite{Buchmuller:2004eg}, supersymmetry breaking
\cite{Buchmuller:2005ma} and gauge coupling unification
\cite{Lee:2006sw} have been studied.  An alternative SO(10) model in
five and six dimensions has previously been studied in
\cite{Kim:2004vk}. For a recent discussion of CP violation in a 5D
orbifold GUT model, see~\cite{Bhattacharyya:2007hw}.

An important ingredient of orbifold GUTs is the presence of split bulk
multiplets whose mixings with complete GUT multiplets, localised at
the fixed points, can significantly modify ordinary GUT mass
relations.  This extends the known mechanism of mixing with vectorlike
multiplets \cite{Barr:1979xt,Nomura:1998gm,Asaka:2003fp}. 
Such models have a large mixing of left-handed leptons and right-handed
down quarks, while small mixings of the left-handed down quarks.  
In this way large mixings in the leptonic charged current are 
naturally reconciled with small CKM mixings in the quark current.

Our model of quark and lepton masses and mixings relates
different orders of magnitude whereas factors $\mathcal{O}(1)$ remain
undetermined.  Hence, we can only discuss qualitative features of
quark and lepton mass matrices.  Recently, orbifold compactifications
of the heterotic string have been constructed which can account for
the standard model in four dimensions and which have a six-dimensional
GUT structure as intermediate step very similar to familiar orbifold
GUT models \cite{Buchmuller:2005jr,Lebedev:2006kn,Buchmuller:2007qf}.
In such models the currently unknown $\mathcal{O}(1)$ factors are in
principle calculable, which would then allow for quantitative
predictions.

The goal of the present paper is twofold: As a typical example, we
first study the model \cite{Asaka:2003iy} in more detail and
explicitly compute the mass eigenstates, masses and mixing angles.
Second, we investigate the question of CP violation, both in the quark
and lepton sector and possible connections between the two.
In previous studies, CP violation has mostly been neglected assuming
that, barring fortunate cancellations, the phases and mixings are
practically independent.  Nevertheless this question and the flavour
structure are strongly interconnected, and we will see that a specific
pattern of mass matrices can give a distinct signature also in the CP
violation invariants.

This paper is organised as follows: In Section~\ref{se:model} we
describe the 6D orbifold GUT model and the diagonalisation of the mass
matrices defining the low energy SM fermions.  In Section~\ref{se:quark}
we discuss the CP violation in the quark sector, whereas
Section~\ref{se:lepton} is devoted to the CP violation in the leptonic
sector.  Conclusions are given in Section~\ref{se:conclusions}.  Two
appendices provide details to the computation of the mass eigenstates
and CP violation in extensions of the SM.

\section{SO(10) Unification in six dimensions}\label{se:model}

\begin{figure}
  \centering
  \includegraphics[width=0.45\linewidth]{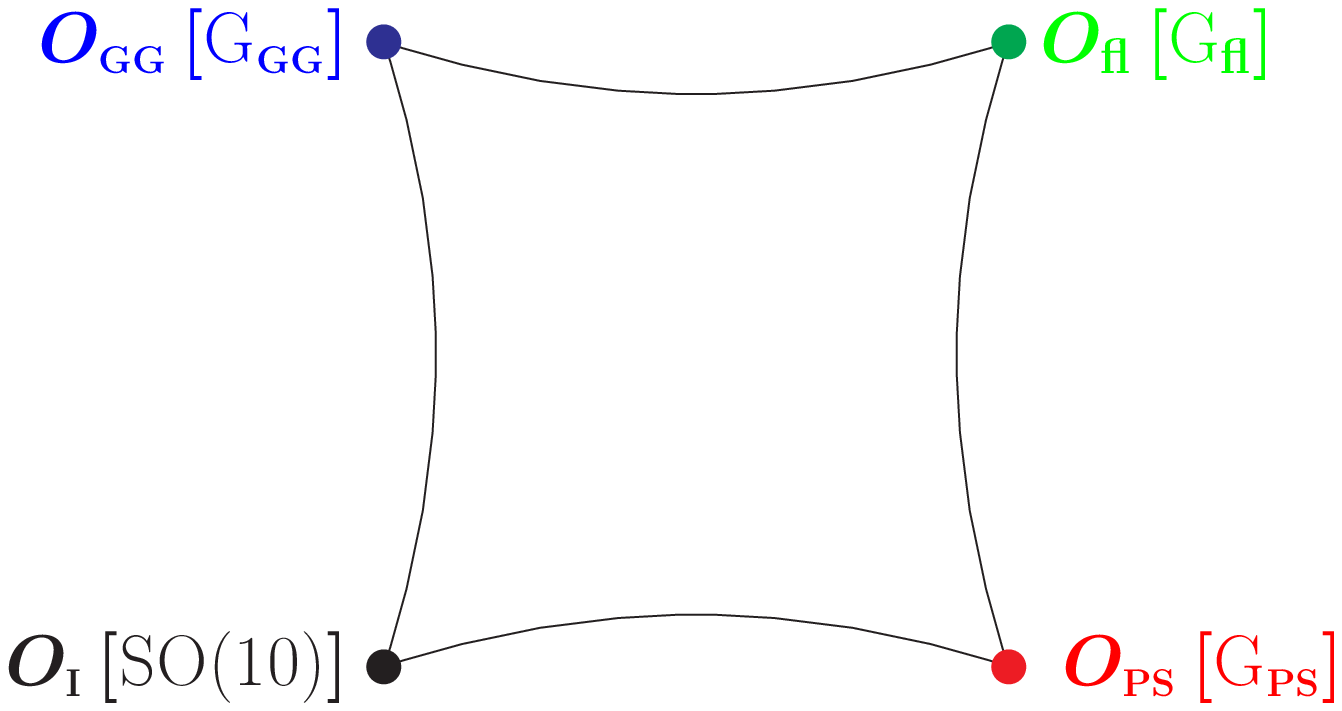}
  \caption{The three SO(10) subgroups at the corresponding fixed
    points (branes) of the orbifold
    \mbox{$\mathbbm{T}^2/(\mathbbm{Z}^I_2 \times
      \mathbbm{Z}^\text{PS}_2 \times \mathbbm{Z}^\text{GG}_2)$}.}
  \label{fig:orb}
\end{figure}

We study an SO(10) GUT model in 6D with $N=1$ supersymmetry
compactified on the orbifold $\mathbbm{T}^2/(\mathbbm{Z}^\text{I}_2
\times \mathbbm{Z}^\text{PS}_2 \times \mathbbm{Z}^\text{GG}_2)$
\cite{Asaka:2001eh,Hall:2001xr}.  The theory has four fixed points,
$O_\text{I}$, $O_\text{PS}$, $O_\text{GG}$ and $O_\text{fl}$, located
at the four corners of a `pillow' corresponding to the two compact
dimensions (cf.~Fig.~\ref{fig:orb}).  The extended supersymmetry is
broken at all fixed points; in addition, the gauge group SO(10) is
broken to its three subgroups $\text{G}_\text{PS} = \text{SU(4)}
\times \text{SU(2)} \times \text{SU(2)}$; $\text{G}_\text{GG} =
\text{SU(5)} \times \text{U(1)}_X$; and flipped SU(5),
$\text{G}_\text{fl} = \text{SU(5)}' \times \text{U(1)}^\prime$, at
$O_\text{PS}$, $O_\text{GG}$ and $O_\text{fl}$, respectively.  The
intersection of all these GUT groups yields the standard model group
with an additional U(1) factor, $\text{G}_{\text{SM}^\prime} =
\text{SU(3)} \times \text{SU(2)} \times \text{U(1)}_Y \times
\text{U(1)}_{Y^\prime}$, as unbroken gauge symmetry below the
compactification scale.

The field content of the theory is strongly constrained by imposing
the cancellation of irreducible bulk and brane anomalies
\cite{Asaka:2002my}.  The model proposed in
Ref.~\cite{Asaka:2003iy} contains three spinors
$\psi_i(\mathsf{16})$, $i=1\ldots 3$, as brane fields as well as six
vectorial fields $H_j(\mathsf{10})$, $j=1\ldots 6$, and two pairs of
spinors, $\Phi(\mathsf{16}) + \Phi^c(\ovl{\mathsf{16}})$ and
$\phi(\mathsf{16}) + \phi^c(\ovl{\mathsf{16}})$ as bulk
hypermultiplets.

The massless zero modes $N(\Phi)$ and $N^c(\Phi^c)$ acquire vacuum
expectation values (vevs), $v_N = \langle N \rangle = \langle N^c
\rangle$, breaking $B-L$ and thus $\text{G}_{\text{SM}^\prime}$ to
$\text{G}_\text{SM}$.  The breaking scale is close to the
compactification scale so that $v_N^2/M_\ast \sim 10^{14}$ GeV, where
$M_\ast$ is the cutoff of the 6D theory.  At the weak scale, the
doublets $H_d(H_1)$ and $H_u(H_2)$ acquire vevs, $v_1 = \langle H_d
\rangle$ and $v_2 = \langle H_u \rangle$, breaking the electroweak
symmetry.

The three sequential {\sf 16}-plets are located on the three branes
where SO(10) is broken to its three GUT subgroups; in particular, we
place $\psi_1$ at $O_\text{GG}$, $\psi_2$ at $O_\text{fl}$ and
$\psi_3$ at $O_\text{PS}$.  The parities of $H_5$, $H_6$, $\phi$, and
$\phi^c$ are chosen such that their zero modes,
\begin{align}
  L(\phi) & = 
  \begin{pmatrix}
    \nu_4 \cr e_4 
  \end{pmatrix}
  , & L^c(\phi^c) & =
  \begin{pmatrix}
    \nu^c_4 \cr e^c_4
  \end{pmatrix}
  , & & d^c_4(H_5)\;, & & d_4(H_6)\;,
\end{align}
have the quantum numbers of a lepton doublet and antidoublet as well
as anti-down and down-quark singlets, respectively.  Both
$L(\phi)$ and $L^c(\phi^c)$ are $\text{SU(2)}_L$ doublets.  Together
these zero modes act as a fourth vectorial generation of down quarks
and leptons.

The three `families' $\psi_i$ are separated by distances large
compared to the cutoff scale $M_*$.  Hence, they can only have
diagonal Yukawa couplings with the bulk Higgs fields; direct mixings
are exponentially suppressed.  The brane fields, however, can mix with
the bulk zero modes for which we expect no suppression.  These mixings
take place only among left-handed leptons and right-handed down
quarks, leading to a characteristic pattern of mass matrices
\cite{Asaka:2003iy,Buchmuller:2004eg}.

The mass terms assume the characteristic form,
\begin{align}
  W & = u_i m^u_i u^c_i + d_\alpha m^d_{\alpha\beta} d^c_\beta +
  e^c_\alpha m^e_{\alpha\beta} e_\beta + \nu^c_\alpha
  m^D_{\alpha\beta} \nu_\beta + \tfrac{1}{2}\, \nu^c_i m^N_i \nu^c_i \
  ,
\end{align}
where latin indices only span $1,2,3$, while greak indices include the
forth generation states.  The up quark and Majorana neutrino mass
matrices, $m^u$ and $m^N$, are diagonal $3\times 3$
matrices,
\begin{subequations}  \label{eq:mass-matrices}
  \begin{align}
    m^u & =
    \begin{pmatrix}
      h^u_{11}v_2 & 0 & 0 \cr 0 & h^u_{22}v_2 & 0 \cr 0 & 0 &
      h^u_{33}v_2
    \end{pmatrix}
    , & m^N & =
    \begin{pmatrix}
      h^N_{11}\frac{v_N^2}{ M_*} & 0 & 0 \cr 
      0 & h^N_{22}\frac{v_N^2}{ M_*} & 0 \cr 
      0 & 0 & h^N_{33}\frac{v_N^2}{ M_*}
    \end{pmatrix}
    .
    \label{eq:mass-matrices-mN}
  \end{align}
  Since $\nu^c_4$ is part of an $\text{SU(2)}_L$ doublet, it cannot
  couple to the other SM singlets in $\psi_i$ via the $B-L$ breaking
  field.  Furthermore, there is no other coupling giving it a direct
  Majorana mass.
  
  The Dirac mass matrices of down quarks, charged leptons and
  neutrinos, $m^d$, $m^e$ and $m^D$, respectively, are $4\times 4$
  matrices instead, due to the mixing with the bulk field zero modes,
  {\allowdisplaybreaks
    \begin{align}
      \label{eq:mass-matrices-md}
      m^d & =
      \begin{pmatrix}
        h^d_{11}v_1 & 0 & 0 & g_1^d {v_N\over M_*} v_1 \cr
        0 & h^d_{22}v_1 & 0 & g_2^d {v_N\over M_*} v_1 \cr
        0 & 0 & h^d_{33}v_1 & g_3^d {v_N\over M_*} v_1 \cr
        f_1 v_N & f_2 v_N & f_3 v_N & M^d
      \end{pmatrix}
      ,
      \\[2mm]
      \label{eq:mass-matrices-lepton}
      m^e & = 
      \begin{pmatrix}
        h^d_{11}v_1 & 0 & 0 & h_{14}^e v_1 \cr
        0 & h^e_{22}v_1 & 0 & h_{24}^e v_1 \cr
        0 & 0 & h^d_{33}v_1 & h_{34}^e v_1 \cr
        M^l_1 & M^l_2 & M^l_3 & M^l_4
      \end{pmatrix}
      , & m^D & =
      \begin{pmatrix}
        h^D_{11}v_2 & 0 & 0 & h_{14}^D v_2 \cr 
        0 & h^u_{22}v_2 & 0 & h_{24}^D v_2 \cr
        0 & 0 & h^u_{33}v_2 & h_{34}^D v_2 \cr 
        M^l_1 & M^l_2 & M^l_3 & M^l_4
      \end{pmatrix}
      ,
    \end{align}
  }%
\end{subequations}
up to corrections $\mathcal{O}(v_N^2/M_*^2)$.  The diagonal elements
satisfy four GUT relations which correspond only to the local unbroken
groups, i.e., SU(5), flipped SU(5) and Pati-Salam subgroups of SO(10).
The hypothesis of a universal strength of Yukawa couplings at each
fixpoint leads to the identification of the diagonal and off-diagonal
elements of $m^u/\tan\beta$, $m^d$, $m^e$, and $m^D/\tan\beta$, where
$\tan\beta=v_2/v_1$, up to coefficients of order one.  This implies an
approximate top-bottom unification with large $\tan{\beta}$ and a
parametrisation of quark and lepton mass hierarchies in terms of the
six parameters $\mu_i$ and $\tm_i$.

The crucial feature of the matrices $m^d$, $m^e$ and $m^D$ are the
mixings between the six brane states and the two bulk states.  The
first three rows of the matrices are proportional to the electroweak
scale. The corresponding Yukawa couplings have to be hierarchical in
order to obtain a realistic spectrum of quark and lepton masses.  This
corresponds to different strengths of the Yukawa couplings at the
different fixed points of the orbifold. The fourth row, proportional
to $M^d$, $M^l$ and $v_N$, is of order the unification scale and, we
assume, non-hierarchical.

The mass matrices $m^d$, $m^e$ and $m^D$ are of the
common form
\begin{align}
  \label{m4-form}
  m = 
  \begin{pmatrix}
    \mu_1 & 0 & 0 & \tm_1 \cr 0 & \mu_2 & 0 & \tm_2 \cr 0 & 0 & \mu_3
    & \tm_3 \cr \tM_1 & \tM_2 & \tM_3 & \tM_4
  \end{pmatrix}
  ,
\end{align}
where $\mu_i,\tm_i = {\cal O}(v_{1,2})$ and $\tM_i =
\mathcal{O}(M_\text{GUT})$.
This matrix can be diagonalised using the unitary matrices
\begin{align}
  m = U_4 U_3 D\, V_3^\dagger V_4^\dagger
\end{align}
where the matrices $U_4$ and $V_4$ single out the heavy mass
eigenstate, that can then be integrated away, while $U_3$ and $V_3$
act only on the SM flavour indices and perform the final
diagonalisation also in the $3\times 3$ subspace.  The explicit
expressions for the mixing matrices and the mass eigenstates are given
in Appendix~\ref{appendix-mass}.

The parameters in the matrix Eq.~(\ref{m4-form}) are generally
complex; however, we can absorb seven phases with appropriate field
redefinitions and choose the remaining three physical phases to be
contained into the diagonal parameters $\mu_i$,
\begin{align} 
  \label{eq:m4-complex}
  m =
  \begin{pmatrix} 
    \abs{\mu_1} e^{i\theta_1} & 0 & 0 & \tm_1 \cr 
    0 & \abs{\mu_2} e^{i\theta_2} & 0 & \tm_2 \cr 
    0 & 0 & \abs{\mu_3} e^{i\theta_3} & \tm_3 \cr
    \rule[-0.5mm]{0mm}{5.5mm}\tM_1 & \tM_2 & \tM_3 & \tM_4
  \end{pmatrix} 
  .
\end{align}
This is the maximal number of physical phases for four generations of
Dirac fermions, given as usual by $(n-1)(n-2)/2 $ for $n$ generations,
so our texture above does not reduce the CP violation from the typical
$n=4$ case. We will see that the phases survive in the low energy
parameters, but that only one combination defines the single phase
characteristic of three generations.

With this choice, the matrix $V_4$ is real, while $U_4$ contains
complex parameters; however, the imaginary part is suppressed by
$\abs{\mu_i}/\tM $ so that their effect on the low energy CP violation
is negligible as long as the mass of the heavy eigenstate is large
compared to the electroweak scale.  From the unification of the gauge
couplings, we expect indeed $\tM $ to be of the order of the GUT
scale~\cite{Lee:2006sw}.
Then the discussion of the low energy CP violation, which would in
general be characterised by many CP invariants
\cite{Branco:2001pq,Branco:2004hu}, reduces to the case of three light
generations (see Appendix~\ref{app:invariants}).

The effective mass matrix is given by $\widehat{m}$, the $3\times 3$
part of
\begin{align}
  \label{eq:mhat}
  m' & = U_4^{\dagger} m V_4 =
  \begin{pmatrix}
    \widehat{m} & 0 \cr 0 & \tM
  \end{pmatrix} 
  + \mathcal{O}\left(\frac{v^2}{\tM^2}\right) , \qquad \widehat{m} =
  \begin{pmatrix}
    \mu_1 (V_4)_{1j} + \tm_1 (V_4)_{4j} \cr \mu_2 (V_4)_{2j} + \tm_2
    (V_4)_{4j} \cr \mu_3 (V_4)_{3j} + \tm_3 (V_4)_{4j}
  \end{pmatrix}
  ;
\end{align}
in terms of the parameters in Eq.~(\ref{m4-form}), it reads
\begin{align}
  \widehat{m} & =
  \begin{pmatrix}
    \mu_1 \frac{\tM_4}{\sqrt{\tM_1^2+\tM_4^2}} - \tm_1
    \frac{\tM_1}{\sqrt{\tM_1^2+\tM_4^2}} & 0 & - \mu_1
    \frac{\tM_1\sqrt{\tM_2^2+\tM_3^2}}{\tM\,\sqrt{\tM_1^2+\tM_4^2}} -
    \tm_1
    \frac{\tM_4\sqrt{\tM_2^2+\tM_3^2}}{\tM\,\sqrt{\tM_1^2+\tM_4^2}}
    \cr
    \rule{0mm}{8mm}%
    -\tm_2 \frac{\tM_1}{\sqrt{\tM_1^2+\tM_4^2}} & \mu_2
    \frac{\tM_3}{\sqrt{\tM_2^2+\tM_3^2}} & \mu_2
    \frac{\tM_2\sqrt{\tM_1^2+\tM_4^2}}{\tM\,\sqrt{\tM_2^2+\tM_3^2}} -
    \tm_2
    \frac{\tM_4\sqrt{\tM_2^2+\tM_3^2}}{\tM\,\sqrt{\tM_1^2+\tM_4^2}}
    \cr
    \rule{0mm}{8mm}%
    -\tm_3 \frac{\tM_1}{\sqrt{\tM_1^2+\tM_4^2}} & -\mu_3
    \frac{\tM_2}{\sqrt{\tM_2^2+\tM_3^2}} & \mu_3
    \frac{\tM_3\sqrt{\tM_1^2+\tM_4^2}}{\tM\,\sqrt{\tM_2^2+\tM_3^2}} -
    \tm_3
    \frac{\tM_4\sqrt{\tM_2^2+\tM_3^2}}{\tM\,\sqrt{\tM_1^2+\tM_4^2}}
  \end{pmatrix} 
  .  \nonumber
\end{align}
As any matrix, $\widehat{m}$ can be transformed into upper triangular
form just by basis redefinition on the right,
\begin{align}
  \label{eq:triagonal-matrix}
  \ovl{m} & = \widehat{m}\ \widehat{V}_3 =
  \begin{pmatrix}
    \gamma\bm_1 & \bm_1 & \beta\bm_1 \cr 0 & \bm_2 &
    \alpha\bm_2 \cr 0 & 0 & \bm_3
  \end{pmatrix}
  .
\end{align}
This form is particularly suitable in the case of the down quarks,
where $\widehat{V}_3$ acts on the right-handed quarks and disappears
from the low energy Lagrangian due to the absence of right-handed
current interactions.  Note that we can reshuffle the phases,
reabsorbing three of them into the unitary transformation
$\widehat{V}_3$, but we are still left with three complex parameters.
We can exploit this freedom to obtain real diagonal elements $\bm_2$,
$\bm_3$ and $\gamma\bm_1$, while $\alpha$, $\beta$, and $\bm_1$ remain
complex.

On the other hand, we can still redefine two phases on the left-hand
side, keeping an overall phase free, with a diagonal matrix
\begin{align}
  P_{L3} = \diag \left( e^{-i\zeta_1}, e^{-i\zeta_2}, 1 \right) .
  \label{phaseleft3}
\end{align}
This transformation allows us to shift the phase of $\bm_1$ into
$\gamma$, which will be convenient later in the limit where $\gamma$
vanishes.  Again, such a phase shift does not reduce the number of
complex parameters in the down quark matrix, which remains three.
Moreover, this reparametrisation does not change the CKM matrix, since
the up quark mass matrix is diagonal and so such phase transformation
can be compensated by an identical one for both $u_i$ and $u^c_i$.

The matrix $\widehat{V}_3$ differs from the upper $3\times 3$ part of
the diagonalising matrix $V_3 = \widehat{V}_3 V_3^\prime$; however, they
are very similar in the hierarchical case.  The relation between these
two can be found in Appendix~\ref{appendix-mass}, together with the
general expression for $U_3$, the $3\times 3$ part of which is the CKM
matrix.

For the leptons, it is the matrix $V_4 V_3$ that acts on the
left-handed states, so the mismatch between the charged leptons and
neutrinos (see Eq.~(\ref{eq:mass-matrices-lepton})) basis appears in
the charged current interaction and the definition of the flavour
neutrino eigenstates.  However, the matrix $V_4$ which contains large
mixing angles and rotates away the heavy eigenstate is the same for
charged leptons and neutrinos since the heavy state is an
$\text{SU(2)}_L$ doublet.  Therefore the PMNS matrix will be given
only by the mismatch between the $\widehat{V}_3\simeq V_3$ matrices
for charged leptons and neutrinos.

The complete expressions for the parameters in $\ovl{m}$ are given in
Appendix~\ref{appendix-mass}; in this section, we will only consider
the limit of small $\mu_1$ as well as small $\tm_1$ and/or $\mu_2$.
For $\mu_1=\tm_1=0$, the first row simply vanishes, whereas for
$\mu_1=\mu_2=0$, the two first rows of the mass matrix are aligned
(see Eq.~(\ref{eq:m4-complex})).  Therefore both cases correspond to
vanishing down-quark and electron mass.

Since ${\tm_1}/{\tm_2} $ gives $ V_{us}$, we focus on the case
$\mu_1=\mu_2=0$, where\footnote{As mentioned above, it is instructive
  to choose the basis in which $\bm_1$ is real and the vanishing
  parameter $\gamma\bm_1$ complex.  Then it is obvious that we are
  left with only two complex parameters in $\ovl{m}$, namely $\alpha$
  and $\beta$, containing the same phase.}
\begin{align}
  \label{eq:down-limit}
  \alpha & = \beta = \frac{\tm_2}{\bm_2} \left( \frac{\tm_3}{\bm_3} -
    \frac{\tM_4}{\tM} \frac{\mu_3^\ast \tM_3 + \tm_3 \tM_4}{\bm_3 \tM}
  \right) , & \gamma & = 0 \ , \nonumber
  \\
  \frac{\bm_1}{\tm_1} & = \frac{\bm_2}{\tm_2} =
  \frac{\abs{\mu_3}}{\bm_3} \frac{\sqrt{\tM_1^2+\tM_2^2 }}{\tM} \ .
\end{align}
The eigenvalues of the heavier states are given by
\begin{subequations} 
  \label{eq:down-masses}
  \begin{align}
    \label{eq:down-masses-b}
    m_b^2 &= \bm_3^2 \ ,
    \\
    \label{eq:down-masses-s}
    m_s^2 &= \bm_2^2 + \abs{\bm_1}^2 = \bm_2^2 \left( 1 +
      \frac{\tm_1^2}{\tm_2^2} \right) \sim \bm_2^2 \ , \qquad
    \text{where}\ \frac{\tm_1}{\tm_2} \sim V_{us} \;.
  \end{align}
  In this limit, only one single physical CP violating phase survives,
  even in the $4\times 4$ picture; it is contained in $\mu_3$ and so
  in $\alpha$ and $\beta$ (see Eq.~(\ref{eq:down-limit})).  We will
  see, however, that this single phase is not sufficient to have
  low-energy CP violation.
  
  The down-quark mass is indeed very small, so we will use these
  expressions as the order zero approximation, together with the
  corrections proportional to $\abs{\mu_2}/\bm_2$, which determine the
  masses of the down-quark and the electron.  Our expansion parameter
  will therefore be of the order of the mass ratio of the down and
  strange-quark, $m_d/m_s$.  In fact, for $\abs{\mu_1} \ll \abs{\mu_2}$
  we have at leading order
  \begin{align}
    m_d & = \gamma \bm_1 \simeq \tm_1 \frac{\abs{\mu_2}}{\bm_2}
    \frac{\abs{\mu_3}}{\bm_3} \simeq \abs{\mu_2}
    \frac{\abs{\bm_1}}{\bm_2} \simeq V_{us} \abs{\mu_2} ,
  \end{align}
\end{subequations}
so our expansion parameter is
\begin{align}
  \frac{\abs{\mu_2}}{\bm_2} &\simeq \frac{m_d}{m_s V_{us}} \sim
  0.23\; .
\end{align}

The mass ratio of electron and muon is much smaller than the ratio of
down and strange quark.  This implies $(\mu_2\tm_1/\tm_2^2)_e \ll
(\mu_2\tm_1/\tm_2^2)_d$.  Assuming that the difference is due to the
smallest matrix elements, this indicates $(\mu_2)_e/(\mu_2)_d \ll 1$
and/or $(\tm_1)_e/(\tm_1)_d \ll 1$ for $(\tm_2)_e \simeq (\tm_2)_d$.
This fact can easily be accommodated, as we see in
Eqs.~(\ref{eq:mass-matrices}): the presence of the second generation on
the flipped SU(5) brane leads to different values of $\mu_2$ for the
down quarks and charged leptons and the parameter $\tm_1$ stems from
different couplings in the superpotential.

While we derived the fermion mass matrices (\ref{eq:mass-matrices})
within a specific model, they can also arise in other models, where
additional matter is present at the GUT (or compactification) scale.
Thus we could take these matrices as a starting point for the
following discussion, leaving open the question of their origin.

\section{CP violation in the quark sector}\label{se:quark}

We will first consider the CP violation in the quark sector.  As we
have seen in the previous section, our effective $3\times 3$ down
quark mass matrix contains three phases as a remnant of the original
$4\times 4$ matrix, with the dominant complex element being
$\alpha\bm_2$.  We will now derive the combination of the three
phases, which plays the role of the CKM phase.

To describe CP violation for three generations, as is the case in the
SM, it is convenient to use the Jarlskog invariant
\cite{Jarlskog:1985ht}, $J_q$, which is given by
\begin{align}
  6\,i\, \Delta \mathscr{M}_u^2\, \Delta \mathscr{M}_d^2\, J_q = \tr
  \left[ H_u,H_d \right]^3 & = 6\ \im \tr \left( H_u^2 H_d^2 H_u H_d
  \right) ,
\end{align}
where $H=m\, m^\dagger$ and
\begin{align}
  \label{eq:delta-m2}
  \Delta \mathscr{M}^2 = \left( m_3^2 - m_2^2 \right) \left( m_3^2 -
    m_1^2 \right) \left( m_2^2 - m_1^2 \right) ;
\end{align}
note that $\Delta \mathscr{M}^2$ has mass-dimension six.  In our model,
the up quark mass matrix is diagonal, as is $H_u$.  Then the invariant
strongly simplifies and reads
\begin{align}
  \label{eq:jarlskog}
  J_q & = \frac{\im \left( H_d^{12} H_d^{23} H_d^{31} \right)}{\Delta
    \mathscr{M}_d^2} \ .
\end{align}
It is clear from this expression, that any diagonal phase transformation
of $m$ on the left does not have any effect on the Jarlskog invariant.

As discussed in Appendix~\ref{app:invariants}, we can use the effective
$3\times 3$ mass matrix \linebreak \mbox{$H_d^\text{eff} = \widehat{m}\,
  \widehat{m}^{\dagger} = \ovl{m}\, \ovl{m}^{\dagger}$}.  By means of
Eq.~(\ref{eq:triagonal-matrix}), we obtain
\begin{align} 
  \label{eq:hd-eff}
  H_d^\text{eff} =
  \begin{pmatrix}
    \abs{\bm_1}^2 \left(1+\abs{\beta}^2+\abs{\gamma}^2\right) & \bm_1
    \bm_2 \left(1+\alpha^\ast\beta\right) & \bm_1 \bm_3 \beta \cr
    \bm_1^\ast \bm_2 \left(1+\alpha\beta^\ast\right) & \bm_2^2
    \left(1+\abs{\alpha}^2\right) & \bm_2 \bm_3 \alpha \cr \bm_1^\ast
    \bm_3 \beta^\ast & \bm_2 \bm_3 \alpha^\ast & \bm_3^2
  \end{pmatrix}
  ,
\end{align}
where $\bm_2$ and $\bm_3$ are real parameters, as displayed in
Eqs.~(\ref{eq:triangular-parameters}).  Then we have
\begin{align}
  \label{eq:hd-eff-im}
  \im \left[ \left(H_d^\text{eff}\right)^{12}
    \left(H_d^\text{eff}\right)^{23} \left(H_d^\text{eff}\right)^{31}
  \right] & = \abs{\bm_1}^2 \bm_2^2 \bm_3^2 \im{\alpha\beta^\ast
    \left(1+\alpha^\ast\beta\right)}
  \\
  & = \abs{\bm_1}^2 \bm_2^2 \bm_3^2 \im{\alpha\beta^\ast} \nonumber
  \\
  & = \bm_2 \bm_3^2 \im \left[ \left(\alpha\bm_2\right)
    \left(\beta\bm_1\right)^\ast \bm_1 \right] .  \nonumber
\end{align}
We see that the Jarlskog invariant is always independent of the argument
of $\gamma$ and it vanishes in the limit $\mu_1,\,\tm_1\to 0$ such that
$\bm_1=0$.  As we might expect, $J_q$ vanishes for $\alpha=\beta$ as
well, i.e., in the limit $\mu_1,\,\mu_2\to 0$.

So the presence of a single phase in $\alpha$ is not sufficient to give
CP violation in the low energy: this phase cancels out in the Jarlskog
invariant.  This effect stems from the alignment of the vectors in
flavour space; however, even in the case of vanishing first generation
mass, the corresponding eigenvector does not decouple from the other two
and the mixing matrix does not reduce to the two-generational case.  In
fact, the CKM matrix is given by (see
Appendix~\ref{appendix-mass})\footnote{We can exploit the phase
  transformation $P_{3L}$ (\ref{phaseleft3}) to absorb the phases of
  $\bm_1, \alpha$ and make all elements of the CKM matrix real showing
  explicitly that the CP violation disappears.}
\begin{align} \label{eq:CKMzero}
  V_\text{CKM} \left(m_d=0\right) & \simeq 
 \begin{pmatrix} 
   1 & \frac{\bm_1}{\bm_2} & \frac{\bm_1 \alpha}{\bm_3} \cr
   \rule{0mm}{6mm}%
   - \frac{\bm_1^\ast}{\bm_2} & 1 & \frac{\bm_2 \alpha}{\bm_3} \cr
   \rule{0mm}{5mm}%
   0 & - \frac{\bm_2 \alpha^*}{\bm_3} & 1
  \end{pmatrix}  
  , & U_3 & = 
  \begin{pmatrix}
    V_\text{CKM}^\dagger & 0 \cr 0 & 1
  \end{pmatrix}
  .
\end{align}
Hence, we cannot conclude that the CP effects disappear due to the
reduction of the system to two generations, nor to the mass degeneracy
between quarks. 
Instead the absence of low energy CP violation is caused by the
particular texture of $m$ in exactly the same basis for the
left-handed quark doublet, where the up quark matrix is diagonal.
This feature is similar to the absence of CP violation in 4D SO(10)
constructions, where a single ten-dimensional Higgs field generates
fermionic masses, yielding a trivial CKM matrix.  Note that there is
still some CP violation effect arising from the dominant phase
$\theta_3$ in $\mu_3$, but it is only apparent in the mixings
involving the fourth heavy state.

Now, the down quark is not massless and the real physical case
corresponds to non-zero $\mu_1$, $\mu_2$ and $\tm_1$.  From the up
quark phenomenology, we know that $\mu_1:\mu_2$ is similar to the mass
ratio of up and charm-quark \cite{Asaka:2003iy}; in addition, $\tm_1 :
\tm_2$ is fixed by the Cabibbo angle.  We will therefore focus on the
linear terms in $\mu_2$ and keep $\mu_1\simeq 0$.

As is apparent in Eq.~(\ref{eq:hd-eff-im}), contributions to $J_q$
come from the complex quantities $\alpha\bm_2$, $\beta\bm_1$, and
$\bm_1$; however, $\beta\bm_1$ is independent of $\mu_2$ (see
Eq.~(\ref{eq:triangular-parameters})),
\begin{align}
  \beta\bm_1 & = \tm_1 \left[\frac{\tm_3}{\bm_3} - \frac{\tM_4}{\tM}
    \frac{\tm_3 \tM_4 + \mu_3^\ast \tM_3}{\bm_3\tM} \right] .
\end{align}
The first order terms are
\begin{align}
  \delta(\alpha\bm_2) & = - \mu_2 \frac{\tM_2}{\tM}\, \frac{\tm_3
    \tM_4 + \mu_3^\ast \tM_3}{\bm_3\tM} , \nonumber
  \\[2pt]
  \delta\bm_1 & = \tm_1 \frac{\mu_2^\ast}{\bm_2} \frac{\mu_3}{\bm_3^2}
  \frac{\tM_2}{\tM} \frac{\tm_3\tM_3 - \mu_3^\ast\tM_4}{\bm_3\tM} ,
\end{align}
and the Jarlskog invariant reads
\begin{align} 
  \label{eq:quark-phase-num}
  J_q & = - \frac{\tm_1^2\tm_2^2\tm_3^2}{\Delta \mathscr{M}_d^2}
  \frac{\tM_2\tM_3}{\tM^2} \left[ \left( 1- \frac{\tM_4^2}{\tM^2}\right)
    \im\frac{\mu_3\mu_2^\ast}{\tm_3\tm_2} + \frac{\tM_3\tM_4}{\tM^2}\,
    \frac{\abs{\mu_3}^2}{\tm_3^2}\, \im\frac{\mu_2}{\tm_2} \right] .
\end{align}
We see that $J_q$ vanishes if either $\mu_2$ or $\mu_3$ vanish, so two
complex quantities are needed to obtain CP violation at low energies.

It is instructive to calculate $H_d^\text{eff}$ also from the matrix
$\widehat{m}$, Eq.~(\ref{eq:mhat}).  Here we notice that the
off-diagonal elements of such matrix are relatively simple since we
can exploit the unitarity of the matrix $V_4$, which gives
$\sum_{k=1}^3(V_4)_{ik} (V_4)^\ast_{jk} = \delta_{ij} - (V_4)_{i4}
(V_4)^\ast_{j4} $.  So we have for $i\neq j$
\begin{align} 
  \label{eq:hd-eff-entries}
  \left(H_d^\text{eff}\right)^{ij} & = \tm_i \tm_j \left( 1-a_i
    a_j^\ast \right) , \qquad a_i \equiv \frac{\tm_i \tM_4 + \mu_i
    \tM_i}{\tm_i \tM} \;,
\end{align}
from which we get the simple expression
\begin{align}
  \im \left[ \left(H_d^\text{eff}\right)^{12}
    \left(H_d^\text{eff}\right)^{23} \left(H_d^\text{eff}\right)^{31}
  \right] & = \tm_1^2 \tm_2^2 \tm_3^2 \sum_{\text{cycl.  perm } ijk}
  \left( 1 + \abs{a_i}^2 \right) \im \left( a_j^\ast a_k \right) .
\end{align}
In the limit of vanishing $\mu_i$, we see that $a_i = \tM_4/\tM$; thus
for $\mu_1 =\mu_2 = 0$, the expression simplifies to
\begin{align}
  \im \left[ \left(H_d^\text{eff}\right)^{12}
    \left(H_d^\text{eff}\right)^{23} \left(H_d^\text{eff}\right)^{31}
  \right] & = \tm_1^2 \tm_2^2 \tm_3^2 \left( 1 + \frac{\tM_4^2}{\tM^2}
  \right) \im \left[ \mu_3^\ast \frac{\tM_4}{\tM} + \mu_3
    \frac{\tM_4}{\tM} \right] = 0\ . \nonumber
\end{align}
For $\mu_1=0$ but $\mu_2\not= 0$, we then obtain
\begin{align}
  \im \left[ \left(H_d^\text{eff}\right)^{12}
    \left(H_d^\text{eff}\right)^{23} \left(H_d^\text{eff}\right)^{31}
  \right] & = \tm_1^2 \tm_2^2 \tm_3^2 \frac{\tM_2 \tM_3}{\tM^2} \left[
    \left(1-\frac{\tM_4^2}{\tM^2}\right)
    \im\left(\frac{\mu_3\mu_2^\ast}{\tm_3\tm_2}\right) \right.
  \\[2pt]
  & \mspace{60mu} \left. + \frac{\tM_3\tM_4}{\tM^2}\,
    \frac{\abs{\mu_3}^2}{\tm_3^2}\, \im\left(\frac{\mu_2}{\tm_2}\right)
    - \frac{\tM_2\tM_4}{\tM^2}\, \frac{\abs{\mu_2}^2}{\tm_2^2}\,
    \im\left(\frac{\mu_3}{\tm_3}\right) \right] . \nonumber
\end{align}
The complete expression for $J_q$ is displayed in
Eq.~(\ref{eq:quark-phase-all}); the dominant terms are exactly those
given in Eq.~(\ref{eq:quark-phase-num}).

For degenerate heavy masses $\tM$, the result simplifies to
\begin{align} 
  J_q & = \frac{1}{16} \frac{\tm_1^2\tm_2}{\Delta \mathscr{M}_d^2}
  \left( 3\, \tm_3 \im\left(\mu_3\mu_2^{\ast}\right) + \abs{\mu_3}^2
    \im\left(\mu_2\right) \vphantom{\tM}\right) .
  \label{eq:quark-phase-sim}
\end{align}
Note that the numerical factor, $\frac{1}{16}$, is minimal for
degenerate $\tM$.
Due to the hierarchy of the down quarks, $\Delta \mathscr{M}_d^2
\simeq m_s^2 m_b^4 \simeq \bm_2^2 \bm_3^4$.  So we finally obtain,
substituting the order of magnitude of the parameters, with $\tm_3
\simeq \abs{\mu_3}$,
\begin{align}
  \label{eq:quark-phase-next}
  J_q & \simeq V_{us} \frac{m_d m_s}{m_b^2}\, \frac{1}{4\sqrt{2}}
  \left( 3\, \sin\left(\theta_3-\theta_2\right) + \sin\theta_2 \right)
  \simeq 10^{-5} \left( 3\, \sin\left(\theta_3-\theta_2\right) +
    \sin\theta_2 \right) .
\end{align}
This is the right order of magnitude; the current experimental value
is $J_q=3\times 10^{-5}$ \cite{Yao:2006px}.  From
Eq.~(\ref{eq:quark-phase-next}) we can conclude that a single complex
parameter, with the other two vanishing, is not enough to have
low-energy CP violation in the quark sector and that the CKM phase is
a combination of the high-energy phases $\theta_i$ weighted by mass
hierarchies.  Moreover, maximal phases seem to be needed to give the
large low-energy phase observed.

\section{CP violation in the leptonic sector}\label{se:lepton}

The charged lepton and Dirac neutrino mass matrices can be transformed
like the down quark mass matrix.  The heavy state is an
$\text{SU(2)}_L$ doublet, so $V_4$ singles out the same state for
charged leptons and neutrinos.

The effective $3 \times 3$-matrices read (cf.~Eq.~(\ref{eq:mhat}))
\begin{align} \label{eq:lepton-dirac-masses}
  \widehat{m}^e & =
  \begin{pmatrix}
    \mu_1 (V_4)_{1j} + \tm_1 (V_4)_{4j} \cr \mu_2 (V_4)_{2j} + \tm_2
    (V_4)_{4j} \cr \mu_3 (V_4)_{3j} + \tm_3 (V_4)_{4j}
  \end{pmatrix}
  , & \widehat{m}^D & =
  \begin{pmatrix}
    \rho_1 (V_4)_{1j} + \rt_1 (V_4)_{4j} \cr \rho_2 (V_4)_{2j} + \rt_2
    (V_4)_{4j} \cr \rho_3 (V_4)_{3j} + \rt_3 (V_4)_{4j}
  \end{pmatrix}
  .
\end{align}
Within our model we assume the hierarchical patterns of
$\mu_i$ and $\rho_i$ as well as $\tm_i$ and $\rt_i$ ($i=1..3$) to be
the same as for down quarks. The precise values, however, can be
different since they originate from different Yukawa couplings, see
Eqs.~(\ref{eq:mass-matrices-lepton}).  Again, we choose the couplings
between the brane states, $\mu_i$ and $\rho_i$, complex.  

Although some of the charged lepton and down quark parameters, namely
$\mu_1$ and $\mu_3$, are related by GUT symmetries, the corresponding
phases after the redefinition leading to Eq.~(\ref{eq:m4-complex}) are
completely uncorrelated.  Thus, there is no direct relation between the
CP violation in the leptonic and in the hadronic observables, even
though, barring cancellations, we expect the leptonic CP violation to be
large as well.  Furthermore, we will see that different combinations of
the phases determine the experimental observables.  Thus even if there
were relations between the phases in the quark and lepton sector, these
would not be observable. Some correlations, however, could survive
between charged and neutral leptons.
As in the quark sector, we expect similar suppression for the CP
violation due to the specific mass texture in our model.

The discussion of the charged lepton masses closely follows the
discussion of the down quarks in the previous section.  The parameters
are chosen such that they match the observed hierarchy, as described in
Appendix~\ref{se:pheno}.  The light neutrino masses, however, result
from the seesaw mechanism, since we have heavy Majorana masses for the
right-handed neutrinos.  This Majorana matrix is diagonal, but can have
complex entries (cf.~Eq.~(\ref{eq:mass-matrices-mN})),
\begin{align} 
  m^N =
  \begin{pmatrix}
    M_1 e^{2i \phi_1} & 0 & 0 \cr 0 & M_2 e^{2i \phi_2} & 0 \cr 0 & 0
    & M_3 e^{2i \phi_3}
  \end{pmatrix}
  = e^{2i \phi_3}
  \begin{pmatrix}
    M_1 e^{2i \Delta\phi_{13}} & 0 & 0 \cr 0 & M_2 e^{2i
      \Delta\phi_{23}} & 0 \cr 0 & 0 & M_3
  \end{pmatrix}
  ,
\end{align}
where $\Delta\phi_{ij}=\phi_i-\phi_j$.
Altogether, we have nine independent phases in the lepton sector; in
the limit of small $\mu_1$ and $\rho_1$, they reduce to seven.  Since
neutrinos are Majorana, we have less freedom in the phase reshuffling.
However, except for electroweak breaking effects in $U_4$, the heavy
state is effectively an SU(2)-doublet of Dirac fermions.  This allows
us to absorb some phases in the Dirac mass matrix and reduce the
system to three generations for both charged and neutral leptons at
the same time.  In the following, we will neglect any effect of this
heavy fourth generation doublet and concentrate on the three light
generations including the right-handed neutrinos.  We expect this
approximation to be valid as long as $\tM \sim M_\text{GUT}$ is much
larger than the Majorana masses $M_i$ \cite{Lee:2006sw}.

\subsection{Seesaw Mechanism and Effective Mass
  Matrix}\label{se:seesaw}

In the case of the leptons, neither $\widehat{m}^e$ nor
$\widehat{m}^D$ is diagonal and therefore we will change the basis in
order to simplify the discussion of the CP violation.  Luckily, the
large rotations of type $\widehat{V}_3$, which bring the Dirac
matrices into triangular form, are similar for charged leptons
and neutrinos, thanks to the same hierarchical structure.

To distinguish the flavour of the light neutrinos, we 
first act on the neutrino Dirac mass matrix with exactly the same
$\widehat{V}_3$ that transforms the charged lepton mass matrix 
into the upper triangular form, see Eq.~(\ref{eq:triagonal-matrix}), 
and obtain
\begin{align}  
  \label{eq:neutrino-dirac-triagonal}
  \ovl{m}^D & =
  \begin{pmatrix}
    A \rb_1 & D \rb_1 & \rb_1 \cr B \rb_2 & E \rb_2 & \rb_2 \cr C
    \rb_3 & F \rb_3 & \rb_3
  \end{pmatrix}
  .
\end{align}
At this stage the charged lepton mass matrix is not yet diagonal, but
not very far from it: the complete diagonalisation can be obtained by
applying another nearly diagonal rotation matrix on the right,
corresponding to the mismatch between $V_3$ and $\widehat{V}_3$, and a
CKM-like rotation $U_3$ on the left as described in 
Appendix~\ref{appendix-mass}.  Note that such a transformation from
the left, as $U_4$, in this case acts on the right-handed fields and
leaves both $H=\ovl{m}^\dagger\, \ovl{m}$ and the light neutrino
Majorana mass matrix,
\begin{align}
  \label{eq:Majoranalight}
  m^\nu_\text{eff} & = - \left( m^D \right)^\top \left( m^N
  \right)^{-1} m^D ,
\end{align}
unchanged.  In fact $U_4$ acts in very good approximation as the unity
matrix on $m^N$ up to terms $\mathcal{O}(v^2/\tM^2)$, while $U_3$ just
cancels out.

So apart for the small rotation on the right needed to diagonalise $H$,
which affects the CP violation in the neutrino oscillation only weakly
(see Section~\ref{se:oscillations}), the neutrino masses and mixings can
be obtained from Eq.~(\ref{eq:Majoranalight}), in the form
\begin{align}
  \label{eq:neutrino-light}
  m^\nu_\text{eff} & = - 
  \begin{pmatrix} 
    C^2 \varrho_3 + B^2 \varrho_2 + A^2 \varrho_1 & C F \varrho_3 + B
    E \varrho_2 + A D \varrho_1 & C \varrho_3 + B \varrho_2 + A
    \varrho_1 \cr
    C F \varrho_3 + B E \varrho_2 + A D \varrho_1 & F^2 \varrho_3 +
    E^2 \varrho_2 + D^2 \varrho_1 & F \varrho_3 + E \varrho_2 + D
    \varrho_1 \cr
    C \varrho_3 + B \varrho_2 + A \varrho_1 & F \varrho_3 + E
    \varrho_2 + D \varrho_1 & \varrho_3 + \varrho_2 + \varrho_1
  \end{pmatrix} ,
\end{align}
where $\varrho_i = e^{-2i \phi_i} \rb_i^2/M_i$.
Note that the determinant of the (23)-submatrix of $m^\nu_\text{eff}$
is not of order $\varrho_3^2$; instead it reads $\varrho_3 \varrho_2
\left(F-E\right)^2 + \varrho_3 \varrho_1 \left(F-D\right)^2 +
\varrho_2 \varrho_1 \left(E-D\right)^2$, allowing a large solar mixing
angle \cite{vis98}.

The leading part of the light neutrino mass matrix
(\ref{eq:neutrino-light}) is obtained in the limit
$\mu_1,\,\rho_1 \rightarrow 0$.  From the general expressions
(\ref{eq:rhobar-a-f}) one obtains
\begin{align}
  \rb_1 & = \rt_1 \frac{1}{\bm_3} \frac{1}{\tM^2} \left[ \tm_3
    \tM_{123}^2 - \mu_3^\ast \tM_3 \tM_4 \right] , \nonumber
  \\[3pt]
  \rb_2 & = \frac{1}{\bm_3} \frac{1}{\tM^2} \left\{ \rt_2 \left[ \tm_3
      \tM_{123}^2 - \mu_3^\ast \tM_3 \tM_4 \right] - \rho_2 \tM_2
    \left[ \tm_3 \tM_4 + \mu_3^\ast \tM_3 \right] \right\} , \nonumber
  \\[3pt]
  \rb_3 & = \frac{1}{\bm_3} \frac{1}{\tM^2} \left\{ \rt_3 \left[ \tm_3
      \tM_{123}^2 - \mu_3^\ast \tM_3 \tM_4 \right] - \rho_3 \left[
      \tm_3 \tM_3 \tM_4 - \mu_3^\ast \tM_{124}^2 \right] \right\} ,
  \label{eq:rhobar-sim}
\end{align}
where we have introduced $\tM_{\alpha\beta\gamma} =
\sqrt{\tM_\alpha^2+\tM_\beta^2+\tM_\gamma^2}$.  

In our model, the Dirac neutrino mass matrix has a hierarchical
structure similar to the one of down quarks and charged leptons.  The
three smallest elements, however, have a considerable uncertainty.
Since $m_e \neq m_d$, these elements cannot be equal for the three
matrices. Inspection of $m^d$ suggests for $\rt_1$ the range between
$m_d$ and $m_s V_{us}$; the difference is a factor $\mathcal{O}(1)$.
In the following we shall consider the case of small $\rt_1$.
For large $\rt_1$ so that $\abs{\varrho_1} > \abs{\varrho_3} $, 
in the following discussion we should interchange 
$ \rb_3,\varrho_3 $ with $\rb_1,\varrho_1$ and consider it as the 
dominant scale.

We here assume $\rb_3 : \rb_2 : \rb_1 \sim
\rt_3 : \rt_2 : \rt_1 \sim m_b : m_s : m_d$, which yields
\cite{Asaka:2003iy}
\begin{align}
  \frac{\abs{\varrho_2}}{\abs{\varrho_3}} & \sim
  \frac{\rb_2^2}{\rb_3^2}\frac{M_3}{M_2} \sim \frac{m_s^2}{m_b^2}
  \frac{m_t}{m_c} \sim 0.2\;, &
  \frac{\abs{\varrho_1}}{\abs{\varrho_3}} & \sim
  \frac{\rb_1^2}{\rb_3^2}\frac{M_3}{M_1} \sim \frac{m_d^2}{m_b^2}
  \frac{m_t}{m_u} \sim 0.2\;,
\end{align}
such that $\varrho_1 \sim \varrho_2 < \varrho_3 $.
Hence, in this model, the weak hierarchy in the neutrino sector can be
traced back to the nearly perfect compensation between down and up quark
hierarchies. 

The relation $\varrho_1 \sim \varrho_2$ implies for the two small
neutrino masses $\abs{m_1} \sim \abs{m_2}$ barring cancellations or
small parameters.
As computed in Appendix~\ref{appendix-mass}, the masses at leading
order assuming $ \varrho_3 $ to dominate are given by
\begin{align}
  m_3 & = - \varrho_3 \left(1 + \abs{F}^2 + \abs{C}^2\right) ,
  \nonumber
  \\[3pt]
  \abs{m_2 m_1} & = \abs{\varrho_2\varrho_1} \frac{\abs{(F-E)(A-B) +
      (D-E)(B-C)}^2}{1 + \abs{F}^2 + \abs{C}^2}\ .
  \label{eq:light-masses}
\end{align}
The light neutrino mass spectrum has normal hierarchy, and the ratio
$m_2^2/m_3^2$ can be identified with $\Delta m_\text{sol}^2/\Delta
m_\text{atm}^2$, which is indeed consistent with observations within
the theoretical uncertainties.

The coefficients $A \ldots F$ of the neutrino mass matrix
$m^\nu_\text{eff}$ become in the limit $\mu_1,\,\rho_1 \rightarrow 0$,
\begin{align}
  \label{eq:a-f-sim}
  A & = - \frac{\rt_1}{\rb_1} \frac{\mu_2}{\bm_2} \frac{\mu_3}{\bm_3}
  \frac{\tM_1}{\tM} \ , \nonumber
  \\[3pt]
  B & = \frac{\rho_2 \tm_2 - \rt_2 \mu_2}{\rb_2\bm_2}\,
  \frac{\mu_3}{\bm_3} \frac{\tM_1}{\tM} \ , \nonumber
  \\[3pt]
  C & = \frac{\tm_3 \rho_3 - \mu_3 \rt_3}{\rb_3\bm_3}\,
  \frac{\mu_2}{\bm_2} \frac{\tM_1}{\tM} \ , \nonumber
  \\[3pt]
  D & = \frac{\rt_1}{\rb_1} \frac{1}{\bm_2} \frac{1}{\bm_3^2}
  \frac{1}{\tM^2} \left[ \tm_2 \abs{\mu_3}^2 \tM_{12}^2 + \mu_2^\ast
    \mu_3 \tM_2 \left( \tm_3 \tM_3 - \mu_3^\ast \tM_4 \right) \right]
  , \nonumber
  \\[3pt]
  E & = D + \frac{\rt_1}{\rb_1} \frac{\rho_2}{\rb_2} \frac{1}{\bm_2}
  \frac{1}{\bm_3} \frac{1}{\tM^2} \left[ \tm_2 \mu_3^\ast \tM_2 \tM_3
    + \mu_2^\ast \left( \tm_3 \tM_{13}^2 - \mu_3^\ast \tM_3 \tM_4
    \right) \right] , \nonumber
  \\[3pt]
  F & = \frac{1}{\rb_3} \frac{1}{\bm_2} \frac{1}{\bm_3^2}
  \frac{1}{\tM^2} \left( \rt_3 \mu_3 - \rho_3 \tm_3 \right) \left[
    \tm_2 \mu_3^\ast \tM_{12}^2 + \mu_2^\ast \tM_2 \left( \tm_3 \tM_3
      - \mu_3^\ast \tM_4 \right) \right] .
\end{align}
Note that $B$, $C$, $F$ vanish in the limiting case of equal hierarchy
in the neutrino and charged lepton Dirac mass matrix, i.e., for
$\rho_i/\rt_i = \mu_i/\tm_i$, and $A$ is in this case proportional to
$\gamma \bm_1$.  In fact, if the neutrino and charged lepton vectors are
perfectly aligned in flavour space the neutrino Dirac matrix becomes
triangular at the same time as the charged lepton one and we cannot
reproduce large neutrino mixing.  There is though no reason to expect
such alignment since the parameters $\rt_i$, $\tm_i$ are not related by
any GUT relation, as can be seen in Eq.~(\ref{eq:mass-matrices-lepton}).
So the large neutrino mixing angles are not generated simply by the
large LH rotation contained in the charged lepton's $\widehat{V}_3$, but
from its misalignment with the neutrinos.

Using the relations between $\rt_i$, $\rb_i$ and $\rho_i$, and $\tm_i$,
$\bm_i$ and $\mu_i$ due to the hierarchical structure of the mass
matrices in our model, one obtains the simple expressions,
\begin{align}
  \label{eq:rough}
  A & \sim C \sim \frac{\mu_2}{\bm_2} \ , & B & \sim
  \frac{\rho_2}{\rb_2} - \frac{\mu_2}{\bm_2} \ , & D & \sim E \sim F
  \sim 1 \ .
\end{align}

The mixing angles are computed in Appendix~\ref{se:neutrino}; in the
case the parameters $A$, $C$ are small, they are given by
\begin{align}
  \tan \theta_{23} & \simeq \abs{F} \;, \nonumber
  \\[3pt]
  \tan \theta_{12} & \sim \frac{\abs{B}}{\abs{E-F}} \sqrt{1+\abs{F}^2}
  \;, \nonumber
  \\[3pt]
  \sin\theta_{13} & \sim \frac{C}{\sqrt{1+\abs{F}^2}} +
  \frac{B\left(EF+1\right)}{\left(1+\abs{F}^2\right)^{3/2}}
  \frac{\abs{\varrho_2}}{\abs{\varrho_3}} \;.
  \label{eq:mixing-angles}
\end{align}
The atmospheric mixing angle $\theta_{23}$ is naturally large; the
current best fit~\cite{Yao:2006px,GonzalezGarcia:2007ib} restricts the
parameter $F$ as $0.7\lesssim \abs{F}\lesssim 1.4$ to have it maximal.
Note that $F \geq 0.7 $ can naturally be obtained even for
$\abs{\rho_3}/\rt_3 \sim \abs{\mu_3}/\tm_3$, as discussed in
Appendix~\ref{se:neutrino}.

For $(\mu_2/\bm_2)_e \sim (\mu_2/\bm_2)_d \sim 0.1$ one then obtains
$|C| \sim 0.1$ and a value for $\theta_{13}$ close to the current
upper bound.  In this case though, $\tilde\mu_1^e$ has to be
suppressed with respect to the down quark case in order to give a
consistently small $m_e$.  The large solar mixing $\theta_{12}$ can
then be achieved for $B \sim 0.1-1$ with moderate tuning of $E-F$.

Another possibility is that a very small $\mu_2$ is called for to
explain the smallness of the electron mass.  In this case, we have
naturally $\abs{A},\,\abs{C} \sim 0.01$ and the reactor angle is
dominated by the second term in Eq.~(\ref{eq:mixing-angles}).  Then the
angles $\theta_{12}$ and $\theta_{13}$ depend on the same parameter $B$,
but for the second one there is a suppression by $\varrho_2/\varrho_3$.
So in the case of hierarchical $\varrho_i$, both a large and small angle
can be explained even with relatively large $B $.  Such value for $B$ is
not unnatural, even for small $\mu_2$, if we accept $\rho_2 >
(\mu_2)_e$.  In this case we have $\sin \theta_{13} \lesssim 0.1$
correlated with the mass eigenvalues $m_1 \lesssim m_2 \lesssim m_3 $.
Note that in general, if all parameters $A$, $B$, and $C$ are smaller
than one, we obtain the prediction $m_1 < m_2$, while for $B \sim 1$ the
two lowest eigenvalues are nearly degenerate.

The largest of the heavy neutrino masses is given by $M_3 \sim
m_t^2/\sqrt{\Delta m_\text{atm}^2} \sim 10^{15}\ {\rm GeV}$.  For the
lightest heavy Majorana state the model provides the rough estimate
$M_1 \sim M_3 m_u/m_t \sim 10^{10}\ {\rm GeV}$.

\subsection{Neutrinoless Double Beta Decay
  ($0\nu\beta\beta$)}\label{se:beta-decay}

The simultaneous decay of two neutrons may result in neutrinoless
double beta decay, e.g., ${}^{78}\text{Ge} \to {}^{76}\text{Se} + 2
e$.  This process is currently most promising to prove the Majorana
nature of neutrinos.
The decay width can be expressed as
\begin{align}
  \Gamma & = G \abs{\mathcal{M}^2} \abs{m_{ee}}^2 ,
\end{align}
where $G$ is a phase space factor, $\mathcal{M}$ the nuclear
$0\nu\beta\beta$ matrix element, and $m_{ee}$ is the (11)-element of
the light neutrino mass matrix.

Since the electron mass is very small, the charged lepton mass matrix
in triangular form has nearly a vanishing first row.  Then the
left-handed electron is already singled out; the remaining rotation
mostly affects the (23)-block.
Therefore we can already make an estimate of $m_{ee}$ from the
effective neutrino Majorana matrix, $m^\nu_\text{eff}$.  From
Eq.~(\ref{eq:neutrino-light}), we read off
\begin{align}
  \label{eq:mee}
  \abs{m_{ee}} & = \abs{C^2 \varrho_3 + B^2 \varrho_2 + A^2 \varrho_1}
  \;,
\end{align}
where the last term can be negleted.  This result has the same form as
the standard formula in the case of hierarchical neutrinos
\cite{Petcov:2004wz},
\begin{align}
  \label{eq:mee-compare}
  \abs{m_{ee}} & = \abs{ \sqrt{\Delta m_\text{atm}^2}
    \sin^2\theta_{13}e^{i(\xi_3-\xi_2)} + \sqrt{\Delta
      m_\text{sol}^2}\sin^2\theta_{12}\cos^2\theta_{13}} \;,
\end{align}
where $\xi_3$ and $\xi_2$ are the two Majorana phases in the
conventional parametrization of neutrino mass matrix (\ref{PMNSmatrix}).

We can estimate the size of $\abs{m_{ee}}$ in our model using
\begin{align} 
  \label{eq:neutrino-ev}
  \abs{\rho_3} & \simeq \rt_3 , & \abs{\mu_3} & \simeq \tm_3 , &
  \frac{\rho_2}{\rb_2} & \sim 1 , & \abs{\varrho_3} & \simeq
  \sqrt{\Delta m_\text{atm}^2} , & \abs{\varrho_2} & \simeq \sqrt{\Delta
    m_\text{sol}^2} \ ,
\end{align}
which gives
\begin{align}
  \abs{m_{ee}} \sim \abs{\frac{\mu_2^2}{\bm_2^2} \sqrt{\Delta
      m_\text{atm}^2}\, e^{2i(\phi_2-\phi_3)} + \frac{\rho_2^2}{\rb_2^2}
    \sqrt{\Delta m_\text{sol}^2}} .
\end{align}
Clearly, the last term dominates, yielding the familiar result for
hierarchical neutrinos \mbox{$\abs{m_{ee}} \lsim \sqrt{\Delta
    m_\text{sol}^2} \sim 0.01\ {\rm eV}$ } if $\mu_2/\tm_2 \ll
\rho_2/\rt_2 $.

\subsection{CP Violation in Neutrino
  Oscillations}\label{se:oscillations}

Leptonic CP violation at low energies can be detected via neutrino
oscillations, which are sensitive to the Dirac phase of the light
neutrino mass matrix.  For a diagonal charged lepton mass matrix, the
strength of Dirac-type CP violation is obtained from the invariant
\cite{Branco:2004hu}
\begin{align}
  \label{eq:cp-lepton}
  \tr \left[ h^\nu, h^e \right]^3 & = 6i\, \Delta \mathscr{M}_e^2 \im
  \left[ \left(h^\nu\right)^{12} \left(h^\nu\right)^{23}
    \left(h^\nu\right)^{31} \right] ,
\end{align}
where $h^\nu = \left(m^\nu\right)^\dagger m^\nu$ and $\Delta
\mathscr{M}_e^2$ is the product of the mass squared differences of the
charged leptons, cf.~Eq.~(\ref{eq:delta-m2}).  This quantity is
connected to the leptonic equivalent of the Jarlskog invariant through
\begin{align}
  \label{eq:jl-diagcl}
  J_\ell & = -\frac{1}{\mathscr{M}_\nu^2}\, \im \left[
    \left(h^\nu\right)^{12} \left(h^\nu\right)^{23}
    \left(h^\nu\right)^{31} \right] ,
\end{align}
where
\begin{align}
  \label{eq:delta-nu}
  \Delta \mathscr{M}^2 = \left( m_3^2 - m_2^2 \right) \left( m_3^2 -
    m_1^2 \right) \left( m_2^2 - m_1^2 \right) = \delta m_\text{sol}^2
  \delta m_\text{atm}^4 \sim \abs{\varrho_2}^2\abs{\varrho_3}^4
\end{align}
is now the product of the light neutrino mass squared differences.  In
the standard parametrisation given in Eq.~(\ref{PMNSmatrix}),
\begin{align}
  \label{eq:jl-standard}
  J_\ell & = \im \left[ (V_\nu)_{11} (V_\nu)_{22} (V_\nu)_{12}^\ast
    (V_\nu)_{21}^\ast \right] = \tfrac{1}{8} \cos\theta_{13}
  \sin{2\theta_{13}} \sin{2\theta_{12}} \sin{2\theta_{23}} \sin\delta \ 
  ,
\end{align}
where $\delta$ is the CP violating Dirac phase in the SM with massive
neutrinos.

The expressions (\ref{eq:jl-diagcl}) and (\ref{eq:jl-standard}) assume
that the charged lepton mass matrix is diagonal.  In our case, this
matrix is nearly diagonal after the $\widehat{V}_3$ rotation, as the
electron mass is very small; in fact, the remaining rotation $V_3'$
deviates from a unit matrix only in the 23 sector and at order ${\cal
  O}\left(m_{\mu}^2/m_{\tau}^2 \right) \ll 1$ (see
Eq.~(\ref{eq:v3bis})).  Therefore up to corrections of this order, we
can use Eq.~(\ref{eq:jl-diagcl}) with the effective neutrino mass matrix
$m^\nu_\text{eff}$ given in Eq.~(\ref{eq:neutrino-light}), i.e.,
\begin{align}
  \label{eq:jl-eff}
  J_\ell & = -\frac{1}{\mathscr{M}_\nu^2}\, \im \left[
    \left(h^\nu_\text{eff} \right)^{12} \left(h^\nu_\text{eff}
    \right)^{23} \left(h^\nu_\text{eff} \right)^{31} \right] ,
\end{align}
with now $h^\nu_\text{eff} = \left(m^\nu_\text{eff} \right)^\dagger
m^\nu_\text{eff}$.  We compute the first few terms and obtain
\begin{align}
  \left(h^\nu_\text{eff}\right)^{12}\!\! & = \frac{C^\ast F \;
    \abs{m_3}^2}{1\! + \!\abs{F}^2\!\! + \!\abs{C}^2\!}+ C^\ast E \left(
    1\! + \!F^\ast E\! + \!C^\ast B \right) \varrho_2 \varrho_3^\ast \!
  + \!  B^\ast F \left(1\! + \!F^\ast E\! + \!C^\ast B \right)^\ast
  \varrho_2^\ast \varrho_3 \, , \nonumber
  \\[2pt]
  \left(h^\nu_\text{eff}\right)^{23}\!\! & = \frac{F^\ast\;
    \abs{m_3}^2}{1\! + \!\abs{F}^2\!\! + \!\abs{C}^2\!} + F^\ast \left(
    1\! + \!F^\ast E\! + \!C^\ast B \right) \varrho_2\varrho_3^\ast +
  E^\ast \left( 1\! + \!  F^\ast E\! + \! C^\ast B \right)^\ast
  \varrho_2^\ast \varrho_3 \, , \nonumber
  \\[2pt]
  \left(h^\nu_\text{eff}\right)^{31} \!\! & = \frac{C\; \abs{m_3}^2}{1\!
    + \!\abs{F}^2\! + \!\abs{C}^2} + B \left( 1\! + \! F^\ast E\! + \!
    C^\ast B \right) \varrho_2\varrho_3^\ast + C \left( 1\! + \!F^\ast
    E\! + \! C^\ast B \right)^\ast \varrho_2^\ast \varrho_3 \, .
  \label{eq:hnu-lo}
\end{align}
The leading contribution in the cyclic product of $h^\nu_\text{eff}$,
which is $\propto \abs{m_3}^6$, is real and does not contribute to
$J_\ell$; that is to be expected since it corresponds to the limit of
two massless neutrinos where no physical Dirac phase can be defined.  In
general the first non-trivial terms are of order $\abs{\varrho_3}^4
\abs{\varrho_{2}}^2$, as $\Delta \mathscr{M}^2_\nu$, so that we expect
$\abs{J_\ell} \lsim 1$.  We obtain in fact
\begin{align}
  J_\ell \sim \frac{\left( 1+\abs{B}^2+\abs{E}^2 \right) \left(
      \abs{E}^2-\abs{F}^2+\abs{B}^2-\abs{C}^2 \right)}{\left(
      1+\abs{F}^2+\abs{C}^2 \right)^3} \im\left[ C^\ast F (F-E)^\ast
    (B-C) \right]\, .
\end{align}
Note that the imaginary part vanishes for $E=F$ or $B=C$, when the
flavour eigenvectors are partially aligned.  Furthermore, the
contribution disappears for \mbox{$C=0$}, so it is suppressed by the
small reactor angle as expected.  Due to the unknown parameters
$\mathcal{O}(1)$, no useful upper bound on $J_\ell$ can be derived in
the general case, but we see that the Dirac CP phase is given by a
combination of the phases of the neutrino Dirac mass coefficients $B$,
$C$, $E$ and $F$, derived from the complex parameters $\mu_3$, $\mu_2$,
$\rho_3$, $\rho_2$.  No dependence arises from the heavy neutrino
Majorana phases $\phi_{3,2}$ since they cancel out in
$\abs{\varrho_3}^4\abs{\varrho_2}^2$.

In the limit $\mu_2\to 0$, where $A=C=0$, but with $B$ of order unity,
the dominant contribution to $J_\ell$ comes from higher order terms.  We
can obtain it from
\begin{align}
  \left(h^\nu_\text{eff}\right)^{12} & = B^\ast F \left(1+F^\ast
    E\right)^\ast \varrho_2^\ast \varrho_3 + B^\ast E
  \left(1+\abs{B}^2+\abs{E}^2\right) \abs{\varrho_2}^2 \; , \nonumber
  \\[2pt]
  \left(h^\nu_\text{eff}\right)^{23} & = 
    F^\ast \abs{\varrho_3}^2 \left(1+\abs{F}^2\right)
  + F^\ast
  \left(1+F^\ast E \right) \varrho_2 \varrho_3^\ast + E^\ast
  \left(1+F^\ast E \right)^\ast \varrho_2^\ast \varrho_3 \; , \nonumber
  \\[2pt]
  \left(h^\nu_\text{eff}\right)^{31} & = B \left(1+F^\ast E \right)
  \varrho_2 \varrho_3^\ast + B \left(1+\abs{B}^2+\abs{E}^2\right)
  \abs{\varrho_2}^2 .
  \label{eq:hnu-case-sim}
\end{align}
Note that the leading term, proportional to $\abs{B}^2 \abs{\varrho_3}^4
\abs{\varrho_2}^2$, is real, and in fact we did not have any $\abs{B}^2
$ contributions at that order above.  Hence, we consider the next terms,
\begin{align}
  \left(h^\nu_\text{eff}\right)^{12} \left(h^\nu_\text{eff}\right)^{23}
  \left(h^\nu_\text{eff}\right)^{31} & \propto \left(1+ F^\ast E\right)
  F^\ast \left( \kappa_1 E + \kappa_2 F \right) \varrho_2 \varrho_3^\ast
  \nonumber
  \\
  & \quad + \left(1+ F^\ast E\right)^\ast F \left( \kappa_1 F^\ast +
    \kappa_2 E^\ast \right) \varrho_2^\ast \varrho_3 \ ,
\label{eq:jl1}
\end{align}
where we defined the real parameters
\begin{align}
  \kappa_1 & = \left(1+\abs{B}^2+\abs{E}^2\right)
  \left(1+\abs{F}^2\right) , \nonumber
  \\
  \kappa_2 & = \abs{1+E^\ast F}^2 \; .
\end{align}
Note again that the two terms in Eq.~(\ref{eq:jl1}) are exactly
conjugate to each other for $E=F$ when the two heavy eigenstates are
nearly aligned.  In this limit $\tan\theta_{12}$ becomes maximal.
Therefore, if $B$ gives the dominant contribution, the Dirac type CP
violation is suppressed for maximal solar angle.  The CP invariant
vanishes as well if $B=0$ as the system effectively reduces to two
generations and $\sin\theta_{13}=0$ (recall that we are already in the
limit $A=C=0$).  We then obtain
\begin{align}
  \label{eq:jl-omega}
  \im \left[ \left(h^\nu_\text{eff}\right)^{12}
    \left(h^\nu_\text{eff}\right)^{23}
    \left(h^\nu_\text{eff}\right)^{31} \right] = \abs{B}^2
  \abs{\varrho_2}^2 \abs{\varrho_3}^4 \left( \kappa_1 - \kappa_2 \right)
  \ \im\left(\Omega\right)
\end{align}
with
\begin{align}
  \Omega & = \left(1+EF^\ast\right) F^\ast \left(E-F\right)
  \frac{\varrho_2}{\varrho_3}\ , \nonumber
\end{align}
which yields
\begin{align}
  \label{eq:jl-omega-sim}
  J_\ell & \sim - \abs{B}^2 \left( \kappa_1 - \kappa_2 \right) \ 
  \im\left(\Omega\right) .
\end{align}
Comparison with Eq.~(\ref{eq:jl-standard}) shows then that in this case
the standard Dirac phase $\delta$ is a complicated function of the
phases of $\mu_3$, $\rho_3$, $\rho_2$ in the leptonic Dirac mass
matrices, the difference between two of the Majorana phases
$\Delta\phi_{32}$ and neutrino masses.  It is suppressed by the ratio
$\abs{\varrho_2}/\abs{\varrho_3}$, as is $\sin\theta_{13}$.

Whenever only few of the parameters in the Dirac neutrino mass matrix
matter, we expect correlations between the lightest eigenvalue, the
mixing angles and the maximal value for $J_\ell $.  In
Appendix~\ref{se:neutrino}, we consider the simple case where $B$
dominates and the lightest eigenvalue $m_1$ vanishes; then all the
observables are only function of $B$, $E$, $F$, $\varrho_2/\varrho_3$
and we show relations among them.  In this specific case, even allowing
for the uncertainty on the phases, upper bounds can be obtained for
$\sin\theta_{13}, m_{ee} $ and $ J_\ell$.  In the more general case,
subleading terms and other parameters become important and relax any
such bounds.

\subsection{Leptogenesis}\label{se:leptogenesis}

The out-of-equilibrium decays of heavy Majorana neutrinos is a natural
source of the cosmological matter-antimatter asymmetry
\cite{Fukugita:1986hr}.  In recent years this leptogenesis mechanism has
been studied in great detail.  The main ingredients are CP asymmetry and
washout processes, which depend on neutrino masses and mixings.

It is convenient to work with a diagonal and real matrix for the
right-handed neutrinos, which is obtained from $m^N$ by the phase
transformation
\begin{align}
  P_M = \diag \left( e^{-i\phi_1}, e^{-i\phi_2}, e^{-i\phi_3} \right) .
\end{align}
For hierarchical heavy neutrinos the generated baryon asymmetry is
dominated by decays of the lightest state $N_1$.  In supersymmetric
models the corresponding CP asymmetry is \cite{Covi:1996wh}
\begin{align}
  \varepsilon_1 & = -\frac{3}{8\pi} \sum_i
  \frac{\im\left(\mathsf{M}_{1i}^2\right)}{\mathsf{M}_{11} v_u^2}
  \frac{M_1}{M_i}\;, & \mathsf{M} & = P_M \,
  \widehat{m}^D\widehat{m}^{D\dagger} P_M^\ast \;,
\end{align}
where the matrix elements are given, analogously to
Eq.~(\ref{eq:hd-eff-entries}),
\begin{align}
 \mathsf{M}_{ij} & = e^{i\Delta\phi_{ji}} \rt_i \rt_j \left( 1-b_i
    b_j^\ast \right) , \qquad b_i \equiv \frac{\rt_i \tM_4 + \rho_i
    \tM_i}{\rt_i \tM} \; .
\end{align}
The terms involving one index $1$ simplify for $\rho_1=0$ as
\begin{align}
  \mathsf{M}_{11} & = \rt_1^2 \left( 1-\frac{\tM_4^2}{\tM^2} \right) \;
  , \nonumber
  \\
  \mathsf{M}_{1j} & = e^{i\Delta\phi_{j1}} \rt_1 \rt_j \left( 1-
    \frac{\tM_4}{\tM} \frac{\rt_j \tM_4 + \rho_j^\ast \tM_j}{\rt_j \tM}
  \right) .
\end{align}
The result then reads
\begin{align}
  \varepsilon_1 & \simeq \frac{3}{8\pi} \frac{M_1}{v_u^2} \left(
    1-\frac{\tM_4^2}{\tM^2} \right)^{-1} \sum_{j=2,3}
  \frac{\rt_j^2}{M_j} \eta_j \ ,
\end{align}
where 
\begin{align}
  \eta_j & = - \im \left[ e^{i\Delta\phi_{j1}} \left( 1-
      \frac{\tM_4}{\tM} \frac{\rt_j \tM_4 + \rho_j^\ast \tM_j}{\rt_j
        \tM} \right)^2 \right] .
\end{align}

Since $\rt_2^2 M_3 / (\rt_3^2 M_2) \sim 0.2$, the CP asymmetry is
dominated by the intermediate state $N_3$, i.e., $\varepsilon_1 \simeq
3/(8\pi) M_1 \sqrt{\Delta m_\text{atm}^2}/v_u^2$.  In any case, the
phases involved, $\Delta\phi_{13}, \Delta\phi_{12} $ and the phases of
$\rho_3,\rho_2$, are completely independent of the low-energy CP
violating phase in the quark sector and also not so directly connected
to that in neutrino oscillations (even if they can contribute to it).
For $M_1\sim 10^{10}$\,GeV, one obtains $\varepsilon_1 \sim 10^{-6}$,
with a baryogenesis temperature $T_B \sim M_1 \sim 10^{10}\ \text{GeV}$.
These are typical parameters of thermal leptogenesis
\cite{Buchmuller:1996pa,Buchmuller:1998zf}.

The strength of the washout processes crucially depends on the
effective neutrino mass
\begin{align}
  \widetilde m_1 & = \frac{\mathsf{M}_{11}}{M_1} = \frac{\rt_1^2}{M_1}
  \left( 1 - \frac{\tM_4^2}{\tM^2} \right) \sim \varrho_1 \lsim
  0.01~\text{eV}\;.
\end{align}
With the efficiency factor \cite{Buchmuller:2004nz}
\begin{align}
  \kappa_f & \sim 10^{-2}\left(\frac{0.01~\text{eV}}{\widetilde
      m_1}\right)^{1.1} \sim 10^{-2} \ ,
\end{align}
one obtains for the baryon asymmetry
\begin{align}
  \label{basym}
  \eta_B & \sim 10^{-2} \varepsilon \kappa_f \sim 10^{-8} \kappa_f \sim
  10^{-10}\ ,
\end{align}
consistent with observation.  So for successful leptogenesis we need a
non vanishing $\rt_1,\varrho_1 $ and in particular $\varrho_1 \sim
\varrho_2 $.  In such case a zero neutrino eigenvalue is only possible
due to alignment.

In the above estimate of the baryon asymmetry we have summed over the
lepton flavours in the final state.  In general, the CP asymmetries as
well as the washout processes depend on the lepton flavour, which can
lead to a considerable enhancement of the generated baryon asymmetry
\cite{Abada:2006fw,Nardi:2006fx}.  The neutrino masses $M_1 \sim
10^{10}\ \text{GeV}$, $\widetilde m_1 \sim 0.01\ \text{eV}$ lie in the
`fully flavoured regime' where these effects can indeed be important
\cite{Blanchet:2006ch}.  Hence, depending on the CP violating phases
the generated asymmetry may be significantly larger than the estimate
(\ref{basym}).

\section{Conclusions}\label{se:conclusions}

We have studied in detail a specific pattern of quark and lepton mass
matrices obtained from a six-dimensional GUT model compactified on an
orbifold. Up quarks and right-handed neutrinos have diagonal $3\times 3$
matrices with the same hierarchy whereas down quarks, charged leptons
and Dirac neutrino mass terms are described by $4\times 4$ matrices
which have one large eigenvalue ${\cal O}(M_\text{GUT})$.  The origin of
this pattern are diagonal mass terms for three ordinary quark-lepton
families together with large mixings ${\cal O}(M_\text{GUT})$ with a
pair of SU(5) $({\bf 5 + \bar{5}})$ plets. This vectorial fourth
generation though is made of different split multiplets allowing for a
relaxation of GUT relations.  The six mass parameters of the model in
the quark sector can be fixed by the up and down quark masses.  This
pattern of mass matrices has several remarkable features: The CKM matrix
is correctly predicted and the electron mass is naturally different from
the down quark mass.

The mismatch between down and up quark mass hierarchies leads, via the
seesaw mechanism, to three light neutrino masses with a much milder
hierarchy. Left-handed leptons and right-handed quarks have large
mixings. This leads to large neutrino mixings and to small CKM mixings
of the left-handed down quarks in agreement with observation.

Factors ${\cal O}(1)$ of the mass matrices are unknown, and the
predictive power of the model is therefore limited.  The neutrino
mixings $\sin \theta_{23} \sim 1$ and $\sin \theta_{13} \lesssim 0.1$
are naturally accommodated.  The corresponding neutrino masses are $m_1
\lesssim m_2 \sim \sqrt{\Delta m^2_\text{sol}} < m_3 \sim \sqrt{\Delta
  m^2_\text{atm}}$ and $\abs{m_{ee}} \sim \sqrt{\Delta m_\text{sol}^2}
\lsim 0.01\ {\rm eV}$.

The elements of the mass matrices arise from a large number of different
operators.  Hence, most of the CP violating high-energy phases are
unrelated.  We find that the measured CP violation in the quark sector
can be obtained, even if the CP invariant is suppressed by the alignment
between the two lightest mass eigenstates.  Due to the uncertainties of
${\cal O}(1)$ factors no useful upper bound on the CP violation in
neutrino oscillations is obtained in general.  Some constraints can be
given in the limited case where the number of dominant parameters is
reduced, as it happens if the parameters $A$, $C$ in the neutrino Dirac
mass matrix are suppressed by the smallness of the electron mass.  It is
indeed intriguing that in our setting the smallness of the reactor angle
can be connected to the lightness of the electron.  The model is
consistent with thermal leptogenesis, with a possible enhancement of the
baryon asymmetry by flavour effects.

We conclude that mixings ${\cal O}(M_\text{GUT})$ of three sequential
quark-lepton families with vectorial split multiplets, a pair of lepton
doublets and right-handed down quarks, can account simultaneously for
small quark mixings and large neutrino mixings in the charged weak
current and, correspondingly, for hierarchical quark masses together
with almost degenerate neutrino masses.  The CP phases in the quark
sector, neutrino oscillations and leptogenesis are unrelated.
Quantitative predictions for the lightest neutrino mass $m_1$ and $\sin
\theta_{13}$ require currently unknown ${\cal O}(1)$ factors in more
specific GUT models.

\subsection*{Acknowledgements}

We would like to thank R.~Fleischer and S.~Willenbrock for helpful
discussions.  SW thanks the SLAC Theoretical Physics Group, the DESY
Theory Group, and the CERN Theory Division for hospitality during
various stages of this work.  LC acknowledges the support of the
``Impuls- and Vernetzungsfond'' of the Helmholtz Association, contract
number VH-NG-006.  SW is supported in part by the U.~S.~Department of
Energy under contract No.~DE-FG02-91ER40677.  The work of DEC is
presently supported by a CFTP-FCT UNIT 777 fellowship and through the
projects POCTI/FNU/44409/2002, PDCT/FP/63914/2005, PDCT/FP/63912/2005
(Funda\-\c c\~ao para a Ci\^encia e a Tecnologia -- FCT, Portugal).

\newpage

\begin{appendix}

\section{Mass matrices}\label{appendix-mass}

We will discuss here the mass eigenvalues and the mixing matrices for
the low energy theory in relation to the high energy parameters.

Given a general matrix of the form as in Eq.~(\ref{m4-form}),
\begin{align*}
  m = 
  \begin{pmatrix}
    \mu_1 & 0 & 0 & \tm_1 \cr
    0 & \mu_2 & 0 & \tm_2 \cr
    0 & 0 & \mu_3 & \tm_3 \cr
    \tM_1 & \tM_2 & \tM_3 & \tM_4 
  \end{pmatrix}
  ,
\end{align*}
where $\mu_i,\tm_i = {\cal O}(v_{1,2})$ and $\tM_i = {\cal
  O}(M_\text{GUT})$, the matrices $U_4$ and $V_4$ that single out the
heavy state can be given as~\cite{Buchmuller:2004eg}
\begin{align}
  U_4 & \simeq 
  \begin{pmatrix}
    1 & 0 & 0 & \frac{\mu_1 \tM_1 + \tm_1 \tM_4}{\tM^2} \cr
    0 & 1 & 0 & \frac{\mu_2 \tM_2 + \tm_2 \tM_4}{\tM^2} \cr
    0 & 0 & 1 & \frac{\mu_3 \tM_3 + \tm_3 \tM_4}{\tM^2} \cr
    -{\mu_1 \tM_1 + \tm_1 \tM_4 \over \tM^2} & -{\mu_2 \tM_2 + \tm_2
      \tM_4 \over \tM^2} & -{\mu_3 \tM_3 + \tm_3 \tM_4 \over \tM^2}& 1
  \end{pmatrix}
  ,
  \\[6pt]
  V_4 & =
  \begin{pmatrix} 
    \frac{\tM_4}{\sqrt{\tM_1^2+\tM_4^2}} & 0 &
    -\frac{\tM_1\sqrt{\tM_2^2+\tM_3^2}}{\tM\,\sqrt{\tM_1^2+\tM_4^2}} &
    \frac{\tM_1}{\tM} \cr
    0 & \frac{\tM_3}{\sqrt{\tM_2^2+\tM_3^2}} &
    \frac{\tM_2\sqrt{\tM_1^2+\tM_4^2}}{\tM\,\sqrt{\tM_2^2+\tM_3^2}} &
    \frac{\tM_2}{\tM} \cr
    0 & -\frac{\tM_2}{\sqrt{\tM_2^2+\tM_3^2}} &
    \frac{\tM_3\sqrt{\tM_1^2+\tM_4^2}}{\tM\,\sqrt{\tM_2^2+\tM_3^2}} &
    \frac{\tM_3 }{\tM} \cr
    -\frac{\tM_1}{\sqrt{\tM_1^2+\tM_4^2}} & 0 &
    -\frac{\tM_4\sqrt{\tM_2^2+\tM_3^2}}{\tM\,\sqrt{\tM_1^2+\tM_4^2}} &
    \frac{\tM_4}{\tM}
    \end{pmatrix}
    ,
\end{align}
with $\tM = \sqrt{\sum_i \tM_i^2}$.  In general $V_4$ contains large
mixings, while $U_4$ is approximately the unity matrix, up to terms
$\mathcal{O}\,(v/\tM)$.  Next, $U_3$ and $V_3 = \widehat{V}_3
V_3^\prime$ diagonalise
\begin{align*}
  m' & = U_4^{\dagger} m V_4 =
  \begin{pmatrix}
    \widehat{m} & 0 \cr 0 & \tM
  \end{pmatrix}
  + \mathcal{O}\left(\frac{v^2}{\tM^2}\right) ,
\end{align*}
so both $U_3$ and $V_3$ have a non-trivial $3\times 3$ part only.  In
the following we will use the symbols $U_3,V_3 $ both for the that
non-trivial upper-left corners and the full $4\times 4$ matrices
obtained adding a row and column of zeros and a diagonal $1$ to those.
The effective mass matrix $\widehat{m}$ can be brought into the upper
triangular form by a unitary matrix $\widehat{V}_3 \sim V_3$ such that
\begin{align*}
  \ovl{m} & = \widehat{m}\ \widehat{V}_3 =
  \begin{pmatrix}
    \gamma\bm_1 & \bm_1 & \beta\bm_1 \cr 0 & \bm_2 & \alpha\bm_2 \cr 0
    & 0 & \bm_3
  \end{pmatrix}
  .
\end{align*}
With $v_i = \left( \widehat{m}_{i1}, \widehat{m}_{i2},
  \widehat{m}_{i3} \right)$, the new basis is given by
\begin{align}
  \label{eq:new-basis}
  \vec{e}_3 & = \frac{\vec{v}_3}{\abs{\vec{v}_3}} \ , & \vec{e}_2 & =
  \frac{\vec{v}_2}{\abs{\vec{v}_2}} - \frac{\vec{e}_3^\ast \cdot
    \vec{v}_2}{{\vec{v}_2}} \vec{e}_3 \ , & \vec{e}_1 & = \vec{e}_2
  \times \vec{e}_3 \ .
\end{align}
Note that $V_3$ corresponds to a large angle rotation for the
right-handed quark fields.

While $\bm_3$ and $\bm_2$ are real by construction, we have the
freedom to choose any entry of the first row to be real.  For concrete
calculations, it is convenient to have $\gamma\bm_1$ real or even use
the parameters as given in the basis (\ref{eq:new-basis}); however,
$\gamma\bm_1$ vanishes in the limit $\mu_2\to 0$, so for a general
discussion, it is more appropriate to have $\bm_1$ real. 
Here, we list the entries of $\ovl{m}$ with $\gamma\bm_1$ real in a
general form,
\begin{align}
  \bm_3 & = \abs{v_3} = \sqrt{\abs{\mu_3}^2 + \abs{\tm_3}^2 -
    \tfrac{1}{\tM} \abs{\mu_3\tM_3+\tm_3\tM_4}^2} \ , \nonumber
  \\[3pt]
  \alpha \bm_2 & = \frac{\tm_2\tm_3^\ast}{\bm_3} -
  \frac{\mu_2\tM_2+\tm_2\tM_4}{\tM}\,
  \frac{\mu_3^\ast\tM_3+\tm_3\tM_4}{\bm_3\tM} \ , \nonumber
  \\[3pt]
  \bm_2 & = \sqrt{ \abs{\mu_2}^2 + \abs{\tm_2}^2 - \tfrac{1}{\tM^2}
    \abs{\mu_2\tM_2+\tm_2\tM_4}^2 - \abs{\alpha\bm_2}^2} \ ; \nonumber
  \\[3pt]
  \beta \bm_1 & = \frac{\tm_1\tm_3}{\bm_3} -
  \frac{\mu_1\tM_1+\tm_1\tM_4}{\tM}\,
  \frac{\mu_3^\ast\tM_3+\tm_3\tM_4}{\bm_3\tM} \ , \nonumber
  \\[3pt]
  \bm_1 & = \tm_1 \left(\frac{\tm_2 }{ \bm_2} - \alpha^\ast
    \frac{\tm_3}{\bm_3}\right) - \frac{\mu_1 \tM_1 + \tm_1 \tM_4}{\tM}
  \left[ \frac{\tM_4}{\tM} \left(\frac{\tm_2 }{\bm_2} - \alpha^\ast
      \frac{\tm_3}{\bm_3}\right) + \frac{\mu_2^\ast}{\bm_2}
    \frac{\tM_2}{\tM} - \alpha^\ast \frac{\mu_3^\ast}{\bm_3} \nonumber
    \frac{\tM_3}{\tM} \right]
  \\[3pt]
  \gamma \bm_1 & = \abs{\gamma \bm_1} = \sqrt{\abs{\mu_1}^2 +
    \abs{\tm_1}^2 - \tfrac{1}{\tM^2} \abs{\mu_1\tM_1+\tm_1\tM_4}^2 -
    \abs{\bm_1}^2 \left(1+\abs{\beta}^2\right)} .
  \label{eq:triangular-parameters}
\end{align}
In particular, we find as well the simple expressions
\begin{align}
  \label{eq:triagonal-parameters-alpha-beta}
  \frac{\alpha\bm_2}{\tm_2} - \frac{\beta\bm_1}{\tm_1} & =
  \frac{\tm_3\tM_4 + \mu_3^\ast\tM_3}{\bm_3\tM} \left[
    \frac{\mu_1}{\tm_1} \frac{\tM_1}{\tM} - \frac{\mu_2}{\tm_2}
    \frac{\tM_2}{\tM} \right] \\
  \gamma \bm_1 & = - \tm_1 \frac{\mu_2}{\bm_2} \frac{\mu_3}{\bm_3}
      \frac{\tM_1}{\tM} - \mu_1 \left[ \frac{\tm_2}{\bm_2}
        \frac{\mu_3}{\bm_3} \frac{\tM_2}{\tM} + \frac{\mu_2}{\bm_2}\,
        \frac{\tm_3\tM_3 - \mu_3\tM_4}{\bm_3\tM} \right] ,
 \end{align}
These expressions vanish trivially in the limit $\mu_1,\,\mu_2 \to 0$
and then we obtain the limiting case discussed in Section~\ref{se:model}.
As already discussed in Section~\ref{se:quark}, 
$\beta\bm_1$ is independent of $\mu_2$.

\subsection{Down Quarks and Charged Leptons}\label{se:pheno}

\paragraph{Mass Eigenvalues and Eigenvectors.}

Now take the matrix $\ovl{m}$ as a starting point and compute the
eigenvalues, eigenvectors and mixing matrices.  For making things
simpler, consider for the moment all the parameters as complex, even if
actually $\bm_3$, $\bm_2$, $\gamma\bm_1$, or $\bm_3$, $\bm_2$, $\bm_1$
can be chosen real absorbing the phases into $V_3$.  To compute the
eigenvalues, it is better to consider the hermitian matrices
$\ovl{m}^\dagger \ovl{m}$ or $\ovl{m} \ovl{m}^\dagger$.  The first
option simply gives
\begin{align}
  \ovl{m}^\dagger \ovl{m} & =
  \begin{pmatrix}
    |\bm_1|^2 |\gamma|^2 & |\bm_1|^2 \gamma^\ast & |\bm_1|^2
    \gamma^\ast\beta \cr
    |\bm_1|^2 \gamma & |\bm_2|^2 + |\bm_1|^2 & |\bm_2|^2
    \alpha+|\bm_1|^2 \beta \cr
    |\bm_1|^2 \gamma\beta^\ast & |\bm_2|^2 \alpha^\ast+ |\bm_1|^2
    \beta^\ast & |\bm_3|^2 + |\bm_2|^2 |\alpha|^2+|\bm_1|^2 |\beta|^2
  \end{pmatrix}
  .
\end{align}
Then the determinant is simply
\begin{align}
  \det \left(\ovl{m}^\dagger \ovl{m}\right) = \left| \det
    \left(\ovl{m}\right) \right|^2 = |\gamma|^2 |\bm_1|^2|\bm_2|^2
  |\bm_3|^2
\end{align}
and is only non-vanishing if $\gamma\bm_1 \neq 0$.

The eigenvalue equation is a cubic equation; to obtain the dominant
terms, we expand around $\gamma = 0$.  In this case the equation
reduces to a quadratic one with the solutions
\begin{align}
  \lambda_{2/3} & = \frac{1}{2} \left[ |\bm_3|^2 + |\bm_2|^2
    (1+|\alpha|^2) + |\bm_1|^2 (1+|\beta|^2 )\right]
  \\
  & \quad \pm \frac{1}{2} \sqrt{ \left[ |\bm_3|^2 - |\bm_2|^2
      (1-|\alpha|^2) - |\bm_1|^2 (1-|\beta|^2 )\right]^2 + 4
    \abs{\abs{\bm_2}^2 \alpha + \abs{\bm_1}^2 \beta}^2} \; .  \nonumber
\end{align}
So in this limit, we have eigenvalues at lowest order
\begin{align}
  \lambda_3 & = |\bm_3|^2 + |\bm_2|^2 |\alpha |^2 + |\bm_1|^2 |\beta
  |^2 + \mathcal{O} \left(\frac{\lambda_2^2}{\lambda_3} \right) ,
  \nonumber
  \\
  \lambda_2 & = |\bm_2|^2 + |\bm_1|^2 - \mathcal{O}
  \left(\frac{\lambda_2^2}{\lambda_3} \right) , & \lambda_1 & = 0 \;.
\end{align}
We can also compute the first correction to the zero eigenvalue simply
as
\begin{align}
  \lambda_1 & = \frac{\det (\ovl{m}^\dagger \ovl{m})}{\lambda_2
    \lambda_3} = \frac{|\gamma|^2 |\bm_1|^2|\bm_2|^2 |\bm_3|^2}{
    |\bm_3|^2 (|\bm_2|^2 + |\bm_1|^2 )}
  \simeq \frac{|\gamma|^2 |\bm_1|^2|\bm_2|^2}{ |\bm_2|^2 + |\bm_1|^2}
  \xrightarrow{\ |\bm_1| \ll |\bm_2|\ } |\gamma|^2 |\bm_1|^2 \ .
\end{align}
This means that for vanishing $\mu_1$ we have
\begin{align}
  m_d \simeq |\gamma| |\bm_1| \simeq \frac{|\mu_2 | }{|\bm_2 |}
  |\tm_1| \; .
\end{align}
Using the eigenvalues, we can also solve for the mixing matrices at
lowest order,
\begin{align}
  \label{eq:v3bis}
  V'_3 & =
  \begin{pmatrix}
    1 & 0 & 0 \cr
    0 & 1 & \frac{|\bm_2|^2\alpha +|\bm_1|^2\beta}{|\bm_3|}\cr
    0 & -\frac{|\bm_2|^2\alpha^\ast +|\bm_1|^2\beta^\ast}{|\bm_3|} & 1
  \end{pmatrix}  ,
\end{align}
where we must recall that we had already acted on the mass matrix with a
large angle rotation $\widehat{V}_3$, so the $V'_3$ above is just a
small correction to it.

For the left-handed quark fields, we have instead at leading order
\begin{align} \label{eq:u3bis}
  U_3 & =
  \begin{pmatrix}
    1 & \frac{\bm_1}{\bm_2} & \frac{\bm_1 \beta}{\bm_3} \cr
    - \frac{\bm_1^\ast}{\bm_2^\ast} & 1 & \frac{\bm_2 \alpha}{\bm_3} \cr
    \frac{\bm_1^\ast}{\bm_3^\ast} (\alpha^\ast-\beta^\ast) & -
    \frac{\bm_2^\ast \alpha^\ast}{\bm_3^\ast} & 1
  \end{pmatrix}  
  .
\end{align}
Since the up quark mass matrix is already diagonal, this last mixing
matrix corresponds to the CKM matrix.  From \mbox{$U_3^\dagger\,
  \ovl{m}\, V'_3 = m^\text{diag}$}, we get $V_\text{CKM} = U_3$, so for
$\alpha=\beta$ we have the prediction \mbox{$V_{td} =
  \left(\alpha^\ast-\beta^\ast\right) \bm_1^\ast/\bm_3^\ast = 0$} at
leading order, and the CP violation vanishes!  On the other hand,
$V_{ub}$ has the right order of magnitude as we thought.

\paragraph{Quark Masses and Mixing Angles.}

We can reproduce the observed quark mass eigenvalues and mixing, that
satisfy the relations
\begin{align}
  m_u : m_c : m_t & \simeq \lambda^7 : \lambda^3 : 1 \ , \nonumber
  \\
  m_d : m_s : m_b & \simeq \lambda^4 : \lambda^2 : 1 \ ,
\end{align}
where $\lambda \simeq V_{us} \sim 0.22$ is the Cabibbo angle.
In fact, if we assume
\begin{align}
  \mu_1 : \mu_2 : \mu_3 & \simeq \lambda^7 : \lambda^3 : 1 \
  ,\nonumber
  \\
  \tm_1 : \tm_2 : \tm_3 & \simeq \lambda^3 : \lambda^2 : 1 \ ,
\end{align}
it gives correctly
\begin{align}
  \abs{V_{us}} & \sim \frac{|\bm_1|}{|\bm_2|} \sim
  \frac{|\tm_1|}{|\tm_2|} \sim \lambda \ ,
  \\[2pt]
  \abs{V_{ub}} & \sim \frac{|\bm_1|}{|\bm_3|} \sim
  \frac{|\tm_1|}{|\tm_3|} \sim \lambda^3 \ , &
  \abs{V_{cb}} & \sim \frac{|\bm_2|}{|\bm_3|} \sim
  \frac{|\tm_2|}{|\tm_3|} \sim \lambda^2 \ ; \nonumber
\end{align}
moreover,
\begin{align}
  m_d & \simeq \frac{|\gamma|}{\sqrt{1+|\alpha|^2}} \abs{\bm_1} \simeq
  \frac{|\mu_2|}{|\bm_2|} \frac{|\tm_1|}{\bm_3}\, m_b \simeq \lambda
  \lambda^3 m_b \simeq \lambda^4 m_b \ .
\end{align}
Again $V_{td}$ is suppressed by the difference of $\alpha^\ast
-\beta^\ast \simeq \mu_2/\overline{\mu}_2,\, \mu_1/\overline{\mu}_1$,
as is the Jarlskog invariant, $J_q$.

\paragraph{Low-energy CP violation}
As discussed in the following Appendix, we can express the low-energy CP
violation in the quark section via an effective Jarlskog invariant.  We
calculate this invariant, using Eqs.~(\ref{eq:triangular-parameters}).
The dominant terms are displayed in Eq.~(\ref{eq:quark-phase-num}); the
complete expression reads
\begin{align} 
  \label{eq:quark-phase-all}
  J_q & = \frac{\tm_1^2\tm_2^2\tm_3^2}{\Delta \mathscr{M}_d^2} \left\{
    \frac{\tM_2\tM_3}{\tM^2} \left[ \left( 1 - \frac{\tM_4^2}{\tM^2}
      \right) \im\frac{\mu_3\mu_2^\ast}{\tm_3\tm_2^2} +
      \frac{\tM_3\tM_4}{\tM^2} \frac{\abs{\mu_3}^2}{\tm_3^2}
      \im\frac{\mu_2}{\tm_2} \right] \right.
  \\
  & \mspace{100mu} - \frac{\tM_2^2\tM_3\tM_4}{\tM^4}
  \frac{\abs{\mu_2}^2}{\tm_2^2} \im\frac{\mu_3}{\tm_3} \nonumber
  \\
  & \mspace{100mu} - \frac{\tM_1\tM_3}{\tM^2} \left[ \left( 1 -
      \frac{\tM_4^2}{\tM^2} \right)
    \im\frac{\mu_3\mu_1^\ast}{\tm_3\tm_1} + \frac{\tM_3\tM_4}{\tM^2}
    \frac{\abs{\tm_3}^2}{\tm_3^2} \im\frac{\mu_1}{\tm_1} \right]
  \nonumber
  \\
  & \mspace{100mu} + \frac{\tM_1\tM_2}{\tM^2} \left[ \left( 1 -
      \frac{\tM_4^2}{\tM^2} \right) + \frac{\tM_3^2}{\tM^2}
    \frac{\abs{\mu_3}^2}{\tm_3^2} \right]
  \im\frac{\mu_2\mu_1^\ast}{\tm_2\tm_1} \nonumber
  \\
  & \mspace{100mu} + \frac{\tM_1\tM_2^2}{\tM^3}
  \frac{\abs{\mu_2}^2}{\tm_2^2} \left[ \frac{\tM_4}{\tM}
    \im\frac{\mu_1}{\tm_1} - \frac{\tM_3}{\tM}
    \im\frac{\mu_3\mu_1^\ast}{\tm_3\tm_1} \right] \nonumber
  \\
  & \mspace{100mu} \left. + \frac{\tM_1^2}{\tM^2}
    \frac{\abs{\mu_1}^2}{\tm_1^2} \left[ \frac{\tM_3\tM_4}{\tM^2}
      \im\frac{\mu_3}{\tm_3} - \frac{\tM_2\tM_4}{\tM^2}
      \im\frac{\mu_2}{\tm_2} + \frac{\tM_2\tM_3}{\tM^2}
      \im\frac{\mu_3\mu_2^\ast}{\tm_3\tm_2} \right] \right\} .
  \nonumber
\end{align}

\paragraph{Charged Leptons.}

The charged leptons show a different hierarchy than the down quarks,
we have in fact
\begin{align}
  m_e : m_{\mu} : m_{\tau } & \simeq \lambda^{5-6} : \lambda^2 : 1
  \nonumber 
  \\
  m_d : m_s : m_b & \simeq \lambda^4 : \lambda^2 : 1 \; .
\end{align}
The discrepancy can be solved with a smaller value for $(\mu_2
\tilde\mu_1)_e$, compared to $(\mu_2 \tilde\mu_1)_d$.  As an example,
we choose $\mu_2^e \simeq \lambda^4$ and $\tm_1^e \simeq
\lambda^{3-4}$ such that
\begin{align}
  m_e & \simeq \frac{|\gamma^e|}{\sqrt{1+|\alpha|^2}} |\bm_1| \simeq
  \frac{\left|\mu_2^e\right|}{\left|\bm_2\right|}
  \frac{\left|\tm_1^e\right|}{\bm_3}\, m_{\tau}
  \simeq \lambda^2 \lambda^{3-4}\, m_{\tau } \simeq \lambda^{5-6}\,
  m_{\tau } \ .
\end{align}

Regarding the rotations, the large $V_4$ rotation acts now on the
left-handed fields, but it has to act on both the charged leptons and
the neutrinos, so it has not a large effect in the charged current.
There is, however, an effect coming from the mismatch between
the two $V_3$'s in the charged leptons and neutrino cases.

\subsection{Neutrinos}\label{se:neutrino}

The charged lepton mass matrix is eventually diagonalised via $V_3 =
\widehat{V}_3 V_3^\prime$ and $U_3$ as the down quark matrix.  For the
light neutrino Majorana mass matrix, given by
\begin{align}
  m^\nu_\text{eff} & = - \left( m^D \right)^\top \left( m^N
  \right)^{-1} m^D ,
\end{align}
we can neglect the rotation $U_3$ of the right-handed fields as this
transformation cancels out. $U_4$ does in principle rotate the RH
states, but its effect is suppressed as long as $ M_i < \tilde M $.
Regarding $V_3$, we do not expect it to be the same for both charged and
neutral leptons, so the mismatch between the two provides flavour mixing
in the neutrino sector.

The neutrino Dirac mass matrix can be written after the large rotation
$\widehat{V}_3$ that bring the charged lepton mass matrix into
triangular form as
\begin{align}
  \label{eq:mDbar}
  \ovl{m}^D & = \widehat{m}^D\ \widehat{V}_3 =
  \begin{pmatrix}
    A \rb_1 & D \rb_1 & \rb_1 \cr B \rb_2 & E \rb_2 & \rb_2 \cr C
    \rb_3 & F \rb_3 & \rb_3
  \end{pmatrix}
  ,
\end{align}
where 
{\allowdisplaybreaks
  \begin{align}
    \rb_1 & = \frac{1}{\bm_3} \frac{1}{\tM^2} \left\{ \rt_1 \left[
        \tm_3 \tM_{123}^2 - \mu_3^\ast \tM_3 \tM_4 \right] - \rho_1
      \tM_1 \left[ \tm_3 \tM_4 + \mu_3^\ast \tM_3 \right] \right\} ,
    \nonumber
    \\[3pt]
    \rb_2 & = \frac{1}{\bm_3} \frac{1}{\tM^2} \left\{ \rt_2 \left[
        \tm_3 \tM_{123}^2 - \mu_3^\ast \tM_3 \tM_4 \right] - \rho_2
      \tM_2 \left[ \tm_3 \tM_4 + \mu_3^\ast \tM_3 \right] \right\} ,
    \nonumber
    \\[3pt]
    \rb_3 & = \frac{1}{\bm_3} \frac{1}{\tM^2} \left\{ \rt_3 \left[
        \tm_3 \tM_{123}^2 - \mu_3^\ast \tM_3 \tM_4 \right] - \rho_3
      \left[ \tm_3 \tM_3 \tM_4 - \mu_3^\ast \tM_{124}^2 \right]
    \right\} , \nonumber
    \intertext{and, using the notation $\tM_{\alpha\beta} =
      \sqrt{\tM_\alpha^2+\tM_\beta^2}$,}
    A & = - \frac{1}{\rb_1} \frac{1}{\bm_2} \frac{1}{\bm_3}
    \frac{1}{\tM} \left\{ \rt_1 \mu_2 \mu_3 \tM_1 - \rho_1 \left[
        \tm_2 \mu_3 \tM_2 + \mu_2 \left( \tm_3 \tM_3 - \mu_3 \tM_4
        \right) \right] \right\} , \nonumber
    \\[3pt]
    B & = \frac{\rho_2 \tm_2 - \rt_2 \mu_2}{\rb_2\bm_2}\,
    \frac{\mu_3}{\bm_3} \frac{\tM_1}{\tM} \ , \nonumber
    \\[3pt]
    C & = \frac{\tm_3 \rho_3 - \mu_3 \rt_3}{\rb_3\bm_3}\,
    \frac{\mu_2}{\bm_2} \frac{\tM_1}{\tM} \ , \nonumber
    \\[3pt]
    D & = \frac{1}{\rb_1} \frac{1}{\bm_2} \frac{1}{\bm_3^2}
    \frac{1}{\tM^2} \left\{ \rt_1 \left[ \tm_2 \abs{\mu_3}^2
        \tM_{12}^2 + \mu_2^\ast \mu_3 \tM_2 \left( \tm_3 \tM_3 -
          \mu_3^\ast \tM_4 \right) \right] \right. \nonumber
    \\
    & \mspace{135mu} + \left. \rho_1 \tM_1 \left[ \tm_2 \mu_3^\ast
        \left( \mu_3 \tM_4 - \tm_3 \tM_3 \right) + \mu_2^\ast \tM_2
        \left( \tm_3^2 + \abs{\mu_3}^2 \right) \right] \right\} ,
    \nonumber
    \\
    E & = \frac{1}{\rb_2} \frac{1}{\bm_2} \frac{1}{\bm_3^2}
    \frac{1}{\tM^2} \left\{ \rt_2 \left[ \tm_2 \abs{\mu_3}^2
        \tM_{12}^2 + \mu_2^\ast \mu_3 \tM_2 \left( \tm_3 \tM_3 -
          \mu_3^\ast \tM_4 \right) \right] \right. \nonumber
    \\
    & \mspace{135mu} + \left. \rho_2 \left[ \tm_2 \mu_3^\ast \tM_2
        \left( \tm_3 \tM_3 - \mu_3 \tM_4 \right) \right. \right.
    \nonumber
    \\
    & \mspace{190mu} \left. + \left.  \mu_2^\ast \left( \tm_3^2
          \tM_{13}^2 - 2 \abs{\mu_3} \tm_3 \tM_3 \tM_4 \cos\theta_3 +
          \abs{\mu_3}^2 \tM_{14}^2 \right) \right] \right\} ,
    \nonumber
    \\[3pt]
    F & = \frac{1}{\rb_3} \frac{1}{\bm_2} \frac{1}{\bm_3^2}
    \frac{1}{\tM^2} \left( \rt_3 \mu_3 - \rho_3 \tm_3 \right) \left[
      \tm_2 \mu_3^\ast \tM_{12}^2 + \mu_2^\ast \tM_2 \left( \tm_3
        \tM_3 - \mu_3^\ast \tM_4 \right) \right] .
    \label{eq:rhobar-a-f}
  \end{align}
}

Note that we are here projecting the neutrino flavour states into the
basis defined by the charged leptons as in Eq.~(\ref{eq:new-basis}).  So
we can immediately see that if the neutrino flavour vectors are aligned
with the charged leptons $ B,C, F$ should vanish and the neutrino mass
matrix would become triangular as well.  This corresponds to having
exactly the same hierarchy in the rows of the charged and neutral lepton
Dirac mass matrices, i.e. $\frac{\mu_i}{\tm_i} = \frac{\rho_i}{\rt_i} $.
We do not expect such alignment since the parameters $\rt_i, \tm_i $ are
generated by different operators and not related by any GUT relation, as
can be seen from Eq.~(\ref{eq:mass-matrices-lepton}).  We will consider
in the following the case where the neutrino hierarchies are similar to
those of the down quark matrix, while the charged leptons differ due to
the lighter electron mass. Of course even more involved scenarios are
possible.  In the following we neglect as well corrections coming from
the final diagonalisation, since the entries of $V_3^\prime$ are
suppressed by $\left(\mu_2/\bm_2\right)^2 \lesssim 0.01$.

\paragraph{Mass eigenvalues and eigenvectors.}

We need to compute the eigenvalues of the neutrino mass matrix and the
first step is again to compute the determinant of the matrix
$m_\text{eff}^{\nu}$.  Note that this is a symmetric matrix, but not
real.  Therefore the eigenvalues are in general complex and the matrix
is diagonalised using a unitary matrix $V_{\nu}$ as
\begin{align}
  V_{\nu}^\top m_\text{eff}^{\nu} V_{\nu} = \diag \left( m_1,m_2,m_3
  \right) .
\end{align}
Consider for the moment just the absolute value of the eigenvalues and
then see that we have the relation\footnote{Note that for a $n\times
  n$ mass matrix, the minus sign on the r.h.s. gives a $(-1)^n$
  contribution.}
\begin{align}
  \det \left(m_\text{eff}^{\nu}\right) = - \frac{\left( \det
      \left(m^D\right) \right)^2}{\det \left(m^N\right)} \ .
\end{align}
The last determinant is simply the product of the heavy neutrino
masses, while the first one is given by
\begin{align}
  \det (m^D) = \rb_1 \rb_2 \rb_3 \left[ (F-E)(A-B) +(D-E)(B-C) \right]
  .
\end{align}
In order to have three non-vanishing eigenvalues, we need all $\rb_i
\neq 0$ and at least one of $A$, $B$, and $C$ different from zero.
Also the three vectors corresponding to the rows of the Dirac matrix
must not be aligned with each other.
So we obtain
\begin{align}
  m_1 m_2 m_3 & = - \varrho_1 \varrho_2 \varrho_3 \left[ (F-E)(A-B)
    +(D-E)(B-C) \right]^2 \nonumber
  \\
  & = - \varrho_1 \varrho_2 \varrho_3 \frac{\rt_1}{\rb_1}
  \frac{\rho_2}{\rb_2} \frac{\rho_3}{\rb_3} \frac{1}{\bm_2^2}
  \frac{1}{\bm_3^2} \frac{1}{\tM^2} \frac{\tM_1}{\tM}
  \\
  & \mspace{36mu} \times \left\{ \tm_2^2 \abs{\mu_3}^2 \tM_{12}^2 + 2
    \abs{\mu_2} \tm_2 \abs{\mu_3} \tM_2 \left[ \tm_3 \tM_3
      \cos\left(\theta_2-\theta_3\right) - \mu_3 \tM_4 \cos\theta_2
    \right] \right. \nonumber
  \\
  & \mspace{72mu} + \left. \abs{\mu_2}^2 \left[ \tm_3^2 \tM_{13}^2 - 2
      \abs{\mu_3} \tm_3 \tM_3 \tM_4 \cos\theta_3 + \abs{\mu_3}^2
      \tM_{14}^2 \right] \right\} , \nonumber
\end{align}
for $\rho_1=0$, where $\varrho_i = e^{-2i\phi_i} \rb_i^2/M_i$.

\paragraph{Singling out the heaviest mass eigenstate.}

In the case when $\varrho_3 \gg \varrho_{2,1} $, it is easy to single
out the heaviest eigenstate:
\begin{align}
  (v_{\nu,3}^0)^\top = \frac{1}{\sqrt{1+|F|^2+|C|^2}}
  \left(C^{\ast},F^{\ast},1\right) ,
\end{align}
and the mass eigenvalue to lowest order is given by
\begin{align}
  \label{eq:m3-tree}
  m_3^0 = - \varrho_3 \left(1+\abs{F}^2+\abs{C}^2\right) .
\end{align}
Then up to a rotation in the 12 submatrix, at lowest order the mixing
matrix can be written as
\begin{align}
  V_{\nu}^0 &= 
  \begin{pmatrix}
    \frac{\sqrt{1+\abs{F}^2}}{\sqrt{1+\abs{F}^2+\abs{C}^2}} & 0 &
    \frac{C^\ast}{\sqrt{1+\abs{F}^2+\abs{C}^2}} \cr \frac{-C
      F^\ast}{\sqrt{1+\abs{F}^2+\abs{C}^2}\sqrt{1+\abs{F}^2}} &
    \frac{1}{\sqrt{1+\abs{F}^2}} &
    \frac{F^\ast}{\sqrt{1+\abs{F}^2+\abs{C}^2}} \cr
    \frac{-C}{\sqrt{1+\abs{F}^2+\abs{C}^2}\sqrt{1+\abs{F}^2}} &
    \frac{-F}{\sqrt{1+\abs{F}^2}} &
    \frac{1}{\sqrt{1+\abs{F}^2+\abs{C}^2}}
  \end{pmatrix} 
  ;
  \label{eq:Vnuzero}
\end{align}
this is the basis which gives decoupling of the first eigenstate 
in the limit of vanishing $C$.
From this matrix, we can directly read off the dominant part of the
mixing angles with the heavy eigenstate, $\theta_{23}$ and
$\theta_{13}$.  The charged lepton mass matrix is nearly diagonal, so
we can actually relate with good accuracy the first row to the
electron neutrino flavour.  The left-handed charged lepton flavour 
eigenstates are given as a function of the mass eigenstates by
\begin{align}
  \ell_f = \left( \widehat{V}_3 V_3' \right)^{\dagger} \ell_i
\end{align}
and therefore the neutrino flavour eigenstates correspond to
\begin{align}
  \nu_f = \left(\widehat{V}_3 V_3'\right)^{\dagger} \widehat{V}_3
  V_{\nu} \nu_i = \left(V_3'\right)^{\dagger} V_{\nu} \nu_i \; ,
\end{align}
where $\widehat{V}_3$ cancels out as it acts equally on the whole lepton
doublet; moreover, as we have seen, $V_3'$ is limited to the $23$ corner
and does not modify the electron entry.  We use here the convention
of~\cite{Petcov:2004wz}, and define the PMNS matrix as
\begin{align}
  V_\nu & = 
  \begin{pmatrix}
    c_{13} c_{12} & c_{13} s_{12} & s_{13} \cr -s_{12} c_{23} - c_{12}
    s_{23} s_{13} e^{i\delta} & c_{12} c_{23} - s_{12} s_{23} s_{13}
    e^{i\delta} & c_{13} s_{23} e^{i\delta} \cr s_{12} s_{23} - c_{12}
    c_{23} s_{13} e^{i\delta} & -c_{12} s_{23} - s_{12} c_{23} s_{13}
    e^{i\delta} & c_{13} c_{23} e^{i\delta}
  \end{pmatrix} 
  \begin{pmatrix}
   1 & 0 & 0 \cr 0 & e^{i\xi_2/2} & 0 \cr 0 & 0 & e^{i\xi_3/2}
  \end{pmatrix} 
  \nonumber
  \\
  & = 
  \begin{pmatrix} 
    1 & 0 & 0 \cr 0 & c_{23} & s_{23} \cr 0 & -s_{23} & c_{23}
  \end{pmatrix} 
  \begin{pmatrix}
    c_{13} & 0 & s_{13} e^{-i\delta} \cr 0 & 1 & 0 \cr -s_{13}
    e^{i\delta} & 0 & c_{13}
  \end{pmatrix} 
  \begin{pmatrix}
    c_{12} & s_{12} & 0 \cr -s_{12} & c_{12} & 0 \cr 0 & 0 & 1
  \end{pmatrix} 
  \begin{pmatrix}
    1 & 0 & 0 \cr 0 & e^{i\xi_2/2} & 0 \cr 0 & 0 & e^{i(\delta+\xi_3/2)}
  \end{pmatrix} 
  ,
  \label{PMNSmatrix}
\end{align}
where $c_{ij} = \cos\theta_{ij}$, $s_{ij} =\sin\theta_{ij}$, $\delta$
is the Dirac phase and $\xi_{1,2}$ are the Majorana phases.

So we have at lowest order for $\theta_{13} $ that
\begin{align}
  \label{eq:13-leading}
  (V_{\nu}^0)_{13} = \sin\theta_{13} & \simeq
  \frac{\abs{C}}{\sqrt{1+\abs{F}^2+\abs{C}^2}} \; .
\end{align}
This gives us directly a constraint on the parameter $C$ from the
upper bound on $\abs{\sin\theta_{13}} \leq 0.1$:
\begin{align}
  \label{eq:c-limit}
  \abs{C} \simeq \sqrt{1 + \abs{F}^2+\abs{C}^2} \abs{\sin\theta_{13}}
  \lesssim 0.1 \sqrt{1 + \abs{F}^2}.
\end{align}
Then since the mixing with the first flavour is small, the atmospheric
mixing matrix is given simply by requiring the $23$ corner of the
matrix in Eq.~(\ref{eq:Vnuzero}) to give
\begin{align}
  V_{\text{atm},\,23} =
  \begin{pmatrix}
    \cos\theta_{23} & \sin\theta_{23} e^{-i\xi_{23}} \cr -
    \sin\theta_{23} e^{i\xi_{23}} & \cos\theta_{23}
  \end{pmatrix}
  .
  \label{V23C}
\end{align}

So considering the $23$ sector, we get, again at lowest order,
\begin{align}
  \xi_{23} & = \arg \left(F \right) , \nonumber
  \\
  \tan \theta_{23} & = \abs{F} \; .
\end{align}
To have large mixing angle $\tan 2\theta_{23} \geq 3$
\cite{Yao:2006px,GonzalezGarcia:2007ib}, we must restrict $\abs{F}$
between
\begin{align}
  0.7 \leq \abs{F} \leq 1.4 \;.
\end{align}
Such a value is natural in the case where $\rho_3$, $\rt_3$ and $\mu_3$,
$\tm_3$ are of the same order but not exactly equal, while $\mu_2$ is
small.  Note that even a phase difference can be important.  Assuming
simply $\frac{\rho_3}{\rt_3} = e^{i\omega_3} \frac{\mu_3}{\tm_3}$ and
degenerate $\tM_i$ gives
\begin{align}
  \abs{F} = \frac{2 \sqrt{2(1-\cos\omega_3)}}{3 - \cos\omega_3}\; ,
\end{align}
so we obtain $\abs{F}=1$ for the maximal phase difference $\omega_3 =
\pi$, while $\abs{F} \geq 0.7$ arises in the wide interval $0.26\; \pi
\leq \omega_3 \leq 1.73\; \pi$.  Hence, a nearly maximal atmospheric
angle is natural even for the most simple choice of parameters.  Of
course, more solutions are possible for the general case.

Thus in order to reproduce the observed pattern of mixing parameters,
$C$ has to be small, while $\abs{F}$ is nearly unity.  We can use the
maximal value for $\abs{F}$ and the experimental bound on
$\theta_{13}$ to derive an upper limit on $\abs{C}$,
\begin{align}
  \abs{C} \leq 0.17 \; ,
\end{align}
in agreement e.g. with the ratio $\frac{\mu_2}{\tm_2}$ necessary 
to have a small electron mass.
Note, however, that we can obtain significant
corrections from $\varrho_{2,1} \neq 0$.

\paragraph{Light eigenstates and solar mixing angle.}
The other two eigenvalues and the correction to the heavy mass can be
obtained from the trace and determinant of the matrix
$\left(m_\text{eff}^{\nu}\right)^\dagger m_\text{eff}^{\nu} $, which
can be computed in any basis.  Expanding both the mass matrix and the
eigenvalues to first order,
\begin{align}
  m_\text{eff}^{\nu} &= m_{\varrho_3} + m_{\varrho_{1,2}} \;,
  \nonumber
  \\
  m_3 & = m_3^0 + \delta m_3 \qquad \mbox{while} \quad m_{1,2} =
  \delta m_{1,2} \;,
\end{align}
we have then
\begin{align}
  \delta m_3 = & \frac{ \tr \left[ m_{\varrho_3}^\dagger
      m_{\varrho_{1,2}} \right]}{(m_3^0)^\ast} \ , & \abs{m_1}^2 +
  \abs{m_2}^2 + \abs{\delta m_3}^2 = & \tr \left[
    m_{\varrho_{1,2}}^\dagger m_{\varrho_{1,2}} \right] , \nonumber
  \\[2pt]
  & & \abs{m_2}^2 \abs{m_1}^2 = &
  \frac{\abs{\det(m_\text{eff}^{\nu})}^2}{\abs{m_3}^2} \ .
\end{align}
Choosing the basis appropriately, the relations can be simplified to
give
\begin{align}
  \delta m_3 = & \left( (V_{\nu}^0)^\top m_{\varrho_{1,2}} V_{\nu}^0
  \right)_{33} \; , \nonumber
  \\[2pt]
  \abs{m_1}^2 + \abs{m_2}^2 = & \tr \left[ m_{\varrho_{1,2}}^\dagger
    m_{\varrho_{1,2}} \right] - \abs{\left( (V_{\nu}^0)^\top
      m_{\varrho_{1,2}} V_{\nu}^0\right)_{33}}^2 \; , \nonumber
  \\[2pt]
  \abs{m_2}^2 \abs{m_1}^2 \sim & \abs{\varrho_1 \varrho_2}^2
  \frac{\abs{(F-E)(A-B)
      +(D-E)(B-C)}^2}{\left(1+\abs{F}^2+\abs{C}^2\right)^2}\; .
\end{align}
We will give the result of these expressions for vanishing $C$ and
$\varrho_1 = q \varrho_2$:
\begin{align}
  \delta m_3 = & \varrho_2 \frac{(1-F E)^2 + q (1- F D)^2}{1+\abs{F}^2}
  \; , \nonumber
  \\[3pt]
  \tr \left[m_{\varrho_{1,2}}^\dagger m_{\varrho_{1,2}} \right] = &
  \abs{\varrho_2}^2 \left[ \abs{1+q}^2 + \abs{E^2 + q D^2}^2 + \abs{B^2+
      q A^2}^2 \right. \nonumber
  \\
  & \left. \mspace{50mu} + 2 \abs{BE+ q AD}^2 + 2 \abs{B+ q A}^2 + 2
    \abs{E+q D}^2 \right] , \nonumber
  \\[3pt]
  \abs{m_2}^2 \abs{m_1}^2 \sim & \abs{\varrho_2}^4 \abs{q}^2
  \frac{\abs{A (F-E) + B (D-F)}^4}{\left(1+\abs{F}^2\right)^2}\; .
\end{align}
Then the mass splitting which should generate the solar oscillations
is given by
\begin{align}
  \delta m_\text{sol}^2 & = \sqrt{\left(\abs{m_1}^2+\abs{m_2}^2\right)^2 -
    4 \abs{m_2}^2 \abs{m_1}^2} \nonumber
  \\
  & = \frac{\abs{\varrho_2}^2}{(1+\abs{F}^2)^2} \left\{ \left[ \left(
        1+\abs{F}^2 \right)^2 \left( \abs{1+q}^2 + \abs{E^2 + q D^2}^2
        + \abs{B^2+ q A^2}^2 \right. \right.  \right.  \nonumber
  \\
  & \mspace{225mu} + \left. 2 \abs{BE+ q AD}^2 + 2 \abs{B+ q A}^2 + 2
    \abs{E+q D}^2 \right) \nonumber
  \\
  & \mspace{140mu} - \left. \abs{(1-F E)^2 + q (1- F D)^2}^2 \right]^2
  \nonumber
  \\
  & \mspace{120mu} - \left.  4 \abs{q}^2 \left( 1+\abs{F}^2 \right)^2
    \abs{A (F-E) + B (D-F)}^4 \right\}^{1/2} .
  \label{eq:msol}
\end{align}
So the solar neutrino mass splitting can be matched even in the case
$q=0$ or $A\left(F-E\right)+B\left(D-F\right)=0$, i.e., when the
lightest neutrino is massless.  However, we do not expect the first
limit to be realised, if we assume the same hierarchies between
$\bar\rho_i$ as in the $\bar\mu_i$ in the down quark sector, while for
$M_i$ as the up quark sector.  In that case we have in fact
$\abs{\varrho_2} \sim \abs{\varrho_1}$ and the two lighter eigenvalues
are similar in scale, $m_1 \simeq m_2 \simeq \sqrt{\delta
  m_\text{sol}^2}$.  On the other hand, the determinant could be
suppressed by alignment, i.e., for
$\abs{A\left(F-E\right)+B\left(D-F\right)} \ll 1$, and could give us a
hierarchy also between the two light eigenvalues.

We can then compute the solar mixing angle and the first order
corrections to the $V_{e3} $ mixing parameter.
After rotating with the $V_{\nu}^0$ matrix, we can estimate the solar
angle by using only the $12$ part of the mass matrix; for $C \simeq 0$
the matrix is given by
\begin{align}
  m_{\varrho_{1,2}} (12) & =
  \begin{pmatrix}
    B^2 \varrho_2 + A^2 \varrho_1 & B \varrho_2
    \frac{E-F}{\sqrt{1+\abs{F}^2}} + A \varrho_1
    \frac{D-F}{\sqrt{1+\abs{F}^2}} \cr B \varrho_2
    \frac{E-F}{\sqrt{1+\abs{F}^2}} + A \varrho_1
    \frac{D-F}{\sqrt{1+\abs{F}^2}} & \varrho_2
    \frac{(E-F)^2}{1+\abs{F}^2} + \varrho_1
    \frac{(D-F)^2}{1+\abs{F}^2}
  \end{pmatrix}
  .
\end{align}
Taking the solar mixing matrix as in Eq.~(\ref{V23C}) with
$\theta_{23},\xi_{23} \rightarrow \theta_{12},\xi_{12}$ we obtain
\begin{align}
  e^{-i\xi_{12}} & = \frac{ (m_{\varrho_{1,2}})_{12}
    (m_{\varrho_{1,2}})_{11}^\ast + (m_{\varrho_{1,2}})_{22}
    (m_{\varrho_{1,2}})_{12}^\ast}{ \abs{(m_{\varrho_{1,2}})_{12}
      (m_{\varrho_{1,2}})_{11}^\ast + (m_{\varrho_{1,2}})_{22}
      (m_{\varrho_{1,2}})_{12}^\ast}} \; , \nonumber
  \\[3pt]
  \tan 2\theta_{12} & = \frac{2 \abs{(m_{\varrho_{1,2}})_{12}
      (m_{\varrho_{1,2}})_{11}^\ast + (m_{\varrho_{1,2}})_{22}
      (m_{\varrho_{1,2}})_{12}^\ast}}{ \abs{(m_{\varrho_{1,2}})_{22}}^2
    - \abs{(m_{\varrho_{1,2}})_{11}}^2 } = \frac{2 \sqrt{1+\abs{F}^2}
    \abs{\mathcal{N}}}{\mathcal{D}} \; , \nonumber
  \intertext{where, for $q=\varrho_1/\varrho_2$,}
  \mathcal{N} & = \left[B (E-F) + q A (D-F)\right] \left(B^2 + q A^2
  \right)^\ast \left(1+\abs{F}^2\right) \nonumber
  \\
  & \mspace{90mu} + \left[ \left(E-F\right)^2 + q \left(D-F\right)^2
  \right] \left[B \left(E-F\right) + q A \left(D-F\right) \right]^\ast ,
  \nonumber
  \\
  \mathcal{D} & = \abs{\left(E-F\right)^2 + q \left(D-F\right)^2}^2 -
  \abs{B^2 + q A^2}^2 \left(1+\abs{F}^2\right)^2 .  \nonumber
\end{align}
In order to have a large solar mixing angle, either $A\,q$ or $B$ must
not be small compared to $E-F$ and $D-F$.  But since $A,\, C \propto
\frac{\mu_2}{\tm_2}$, we are led to the case
\begin{align}
  A & = C \sim 0 \; , & B & = \frac{\rho_2}{\rb_2} \frac{\tM_1}{\tM} =
  \mathcal{O}\left(1\right) \ .
\end{align}
Then we can neglect the terms proportional to $A$ and we have simply
\begin{align}
  \tan 2\theta_{12} & = 2\abs{B} \abs{E-F} \sqrt{1+\abs{F}^2}
  \frac{\abs{B}^2 (1+\abs{F}^2) + \abs{E-F}^2 + q (D-F)^2
    \frac{\left(E-F\right)^\ast}{E-F}}{
    \abs{\left(E-F\right)^2+ q \left(D-F\right)^2}^2 -
    \abs{B}^4 \left(1+\abs{F}^2\right)^2} \ .
\end{align}
This formula simplifies further if we neglect the $q\left(D-F\right)$
terms as well.\footnote{Note that taking $A=C=D-F=0$ gives a zero
  determinant for the neutrino mass matrix, so this case applies when
  the lightest eigenvalue is suppressed compared to the solar mass
  scale.}  Then using general trigonometric formulae leads to the
expression in Eq.~(\ref{eq:mixing-angles}),
\begin{align}
  \label{eq:b-constraint}
  \tan\theta_{12} & \simeq \frac{\abs{B}}{\abs{E-F}} \sqrt{1+\abs{F}^2}
  \;.
\end{align}
Taking the experimental value for the solar angle, $\tan^2 \theta_{12}
= 0.45 \pm 0.05 $, gives us for $ \abs{F} \sim 1 $
the range $\abs{B} \sim \left(0.45-0.50\right) \abs{E-F}$.

We can also compute the corrections of order
$\varrho_{1,2}$ to the other two mixing angles, that we have discussed
in the lowest order.  In fact, since $\mu_2 \ll \tm_2$, the
contribution (\ref{eq:13-leading}) is small and the leading
contribution to $\theta_{13}$ comes from the $ B \varrho_2$ term,
\begin{align}
  \label{eq:13-one}
  (V_\nu^{(1)})_{13} = \sin\theta_{13} & \simeq
  \frac{\abs{B\left(EF+1\right)}}{\left(1+\abs{F}^2\right)^{3/2}}
  \frac{\abs{\varrho_2}}{\abs{\varrho_3}} \sim \abs{B} \frac{m_2}{m_3}
  \sim 0.2 \; \abs{B} \ .
\end{align}
So even for vanishing leading order, we expect the first order term to
bring $\theta_{13}$ near to the experimental bound.  Note that it is
the large solar angle that naturally gives $\theta_{13} \sim
\varrho_2/\varrho_3$; in our model it seems pretty difficult to
suppress this angle to much smaller values, apart if there is
a tuned cancellation between zero and first order.

The corrections to the atmospheric angle are of the same order 
$\varrho_2/\varrho_3  $ and do not have a large effect since
we need in any case large parameters in the $23$ sector.
This small shift can in fact be easily compensated by a small 
change in the value of $F$, especially since we do not have
any particular symmetry in the model imposing $F=1$.

\paragraph{Sum Rules for $B $ dominance and vanishing \boldmath{$m_1$}.}

We have seen in the previous paragraph that in case of vanishing $C$,
$A$ and $\varrho_1$, simple expressions can be obtained for all
observables as functions of only few parameters $B$, $E$, $F$ and
$\varrho_{3,2}$.  Then it is possible to obtain relations between the
different observables,
\begin{align}
  \tan\theta_{23} & = \abs{F} \; , \nonumber
  \\[2pt]
  \tan\theta_{12} & = \frac{\abs{B}}{\abs{E-F}} \sqrt{1+\abs{F}^2} \; ,
  \nonumber
  \\[2pt]
  \sin\theta_{13} & =
  \frac{\abs{B\left(EF+1\right)}}{\left(1+\abs{F}^2\right)^{3/2}}
  \frac{\abs{\varrho_2}}{\abs{\varrho_3}} \; , \nonumber
  \\[2pt]
  \frac{\delta m_\text{sol}}{\delta m_\text{atm}} & =
  \frac{\abs{\varrho_2}}{\abs{\varrho_3}} \frac{\sqrt{(1+\abs{F}^2)^2
      \left(1+\abs{E}^2 + \abs{B}^2 \right)^2 - \abs{1-
        FE}^4}}{\left(1+\abs{F}^2\right)^2} \; .
\end{align}
Now we can write the following relation,
\begin{align}
  \frac{\sin\theta_{13}}{\tan\theta_{12}} \frac{\delta
    m_\text{atm}}{\delta m_\text{sol}} & = \frac{\abs{E-F}
    \abs{EF+1}}{\sqrt{\left[ \left(1+\abs{F}^2\right)
        \left(1+\abs{E}^2\right) + \abs{E-F}^2 \tan^2\theta_{12}
      \right]^2 - \abs{1- EF}^4}} \; .
\end{align}
To estimate its value, we can use the fact that $\abs{F} \sim 1$ and
vary only $\abs{E}$ and the phases of $E$, $F$.  We obtain then a
maximal value of the r.h.s. for $EF=1$ so that
\begin{align}
  \sin\theta_{13} & \leq \frac{\delta m_\text{sol}}{\delta m_\text{atm}}
  \frac{\tan\theta_{12}}{1+\tan^2\theta_{12}} \simeq 0.09 \; .
\end{align}
Of course, the angle $\theta_{13}$ can always be reduced by an
appropriate choice of the phases and in particular for $E=F$, so that
there is no lower bound in this type of models.

The effective neutrino Majorana matrix, which is relevant for
neutrinoless double beta decay, simplifies such that
\begin{align}
  \abs{m_{ee}} & = \abs{B}^2 \abs{\varrho_2} \nonumber
  \\
  & = \delta m_\text{sol}\, \frac{\tan^2\theta_{12}
    \abs{E-F}^2}{\sqrt{\left[ \left(1+\abs{F}^2\right)
        \left(1+\abs{E}^2\right) + \tan^2\theta_{12} \abs{E-F}^2
      \right]^2 -\abs{1- FE}^4}}\; .
\end{align}
Again varying the phases and the modulus of $E$, we find the maximal
value for $EF=-1$,
\begin{align}
  \abs{m_{ee}} & \leq \delta m_\text{sol} \,
  \frac{\tan\theta_{12}}{\sqrt{2 + \tan^2\theta_{12}}} \sim 0.43 \;
  \delta m_\text{sol} \; .
\end{align}
Moreover, we can give a simple relation between $m_{ee}$ and the reactor
angle,
\begin{align}
  \frac{\abs{m_{ee}}}{\delta m_\text{atm}} & =
  \frac{\abs{E-F}}{\abs{EF+1}} \sin\theta_{13} \tan \theta_{12} \; .
\end{align}
Note that the singular value for $EF+1=0$ corresponds to a vanishing
reactor angle.

We can even derive a maximal value for the Dirac CP violation for this
case.  From Eqs.~(\ref{eq:jl-eff}) and (\ref{eq:jl-omega}) we get
\begin{align}
  J_\ell & = - \frac{\abs{B}^2 \left(\kappa_1-\kappa_2\right)
    \im\left(\Omega\right)}{(1+\abs{F}^2)^2 \left[(1+\abs{F}^2)^2
      \left(1+\abs{E}^2+\abs{B}^2\right)^2 - \abs{1- EF}^4\right]}
  \\[4pt]
  & = - \frac{\abs{E-F}^4}{1+\abs{F}^2}\;
  \frac{\tan^2\theta_{12} \left(1+\tan^2\theta_{12}\right)
    \im\left(\Omega\right)}{\left[ \left(1+\abs{F}^2\right)
      \left(1+\abs{E}^2\right) + \abs{E-F}^2 \tan^2\theta_{12} \right]^2
    - \abs{1- EF}^4} 
  \nonumber
  \\[4pt]
  & = - \frac{\delta m_\text{sol}}{\delta m_\text{atm}}
  \frac{\abs{E-F}^4}{1+\abs{F}^2}
  \frac{\tan^2\theta_{12} \left(1+\tan^2\theta_{12}\right) \im \left[
      \left(1 + E F^\ast \right) F^\ast \left( E - F \right) e^{i
        \Delta_{23}}\right]}{\left[\left( \left(1\! +\!\abs{F}^2\right)
        \left(1\! +\!\abs{E}^2\right)\! 
         +\!\abs{E\! -\! F}^2\! \tan^2\theta_{12}\right)^2\! 
        -\! \abs{1\! -\! EF}^4 \right]^{3/2}}, \nonumber
\end{align}
where $\Delta_{23}$ is the phase of $\varrho_2/\varrho_3$.  Again, the
prefactor is maximal for $E F = - 1$ and $E=-F$, giving
\begin{align}
  \abs{J_\ell} & \leq \frac{\delta m_\text{sol}}{\delta m_\text{atm}}
  \frac{1+\tan^2\theta_{12}}{2 \tan\theta_{12}
    \left(2+\tan^2\theta_{12}\right)^{3/2}} \abs{\sin\Delta_{23}} \leq
  0.06 \ .
\end{align}
Here the imaginary part is only given by the phase $\Delta_{23}$, but in
more general cases the phases of $E$ and $F$ will play a role as well.
So even for the CP violation in the leptonic sector, the model displays
a suppression given by the ratio of the mass eigenvalues.  Contrary to
the quark case, however, the CP violation is not proportional to the
smallest mass eigenvalue, but it can be non-vanishing even for $m_1 =
0$.

\section{CP Violation and Weak Basis Invariants}\label{app:invariants}

For completeness we discuss here the CP invariants in the case of an
additional vectorial state.  We prove that if the additional state is
much heavier than the electroweak scale, the low energy CP violation can
be expressed by the Jarlskog invariant defined from an effective
$3\times 3 $ down quark mass matrix.

The transformation of a Dirac spinor $\psi(t,\vec{x})$ under parity and
charge conjugation is given by
\begin{align}
  \begin{split}
    \text{P } \psi(t,\vec{x}) \text{ P}^{-1} &= \eta_P\, \gamma^0
    \psi(t,-\vec{x}),
    \\
    \text{C } \psi(t,\vec{x}) \text{ C}^{-1} &= \eta_C\, C
    \bar{\psi}(t,\vec{x})^\top,
  \end{split}
\end{align}
where $\eta_{P,C}$ are non-observable phases.  The matrix $C$ obeys
the relation $\gamma_{\mu}C=-C\gamma^T_{\mu}$.  Since the Lagrangian
is a Lorentz scalar, it only depends on fermionic field bilinears.
Thus, we deduce the CP transformation for such terms,
\begin{align}
  \begin{split} 
    \label{eq:relCP} 
    \text{CP } \bar{\psi}_i \psi_j \left(\text{CP}\right)^{-1} & =
    \bar{\psi}_j \psi_i\ ,
    \\
    \text{CP } \bar{\psi}_i \gamma^5 \psi_j
    \left(\text{CP}\right)^{-1} & = -\bar{\psi}_j \gamma^5 \psi_i\ ,
    \\
    \text{CP } \bar{\psi}_i \gamma^{\mu} \psi_j
    \left(\text{CP}\right)^{-1} & = -\bar{\psi}_j \gamma_{\mu} \psi_i\
    ,
    \\
    \text{CP } \bar{\psi}_i \gamma^{\mu} \gamma^5 \psi_j
    \left(\text{CP}\right)^{-1} & = -\bar{\psi}_j \gamma_{\mu}
    \gamma^5 \psi_i\ .
  \end{split}
\end{align}
Note that the operator $\partial_{\mu}$ transforms under CP as
\mbox{$\partial^{\mu} \rightarrow \partial_{\mu}$}.  

\paragraph{Quark Sector.}

In the Standard Model, it is easy to verify the existence of the CP
symmetry in the Lagrangian, up to mass terms.  In general, the quark
mass terms are CP invariant if and only if it is possible to find a weak
basis transformation which realises
\begin{align}
  \label{eq:mm}
  H^{u\ast} & = W_L H^u W^{\dagger}_L\ , & H^{d\ast} & = W_L H^d
  W^{\dagger}_L\ ,
\end{align} 
where $H^{u,d} = M^{u,d} \left(M^{u,d}\right)^\dagger$.
It follows that
\begin{align}
  W_L \left[H_u,H_d\right] W^{\dagger}_L = - \left[H_u,H_d\right]^\top
  ,
\end{align} 
such that, for $r$ odd,
\begin{align}
  \tr \left[H_u,H_d\right]^r = 0
  \label{CPINV}
\end{align}
is a necessary and sufficient condition for CP invariance
\cite{Bernabeu:1986fc}.

The case of $r=1$ is trivial: the trace of a commutator $[H_u,H_d]$ is
zero.  For $r=3$ and three generations, we have
\begin{align}
  \begin{split}
    I_{\text{SM}}\equiv \tr \left[H_u,H_d\right]^3 = 6i &
    \left(m^2_t-m^2_c\right) \left(m^2_t-m^2_u\right)
    \left(m^2_c-m^2_u\right)
    \\
    & \left(m^2_b-m^2_s\right) \left(m^2_b-m^2_d\right)
    \left(m^2_s-m^2_d\right) J_q \ ,
  \label{eq:CPJ}
  \end{split}
\end{align}
where the quantity $J_q$ does not depend of the mass spectrum, and can
be related, up to a sign, with the CKM matrix, $V$, as
$\abs{J_q}=\abs{\im(V_{12}V^{\ast}_{13}V^{\ast}_{22}V_{23})}$.  We
conclude that in order to have CP violation, we need to have $J_q\ne0$.
This quantity is the lowest weak basis invariant which measure CP
violating effects and it has mass-dimension twelve.  Apart from CP
violation in the strong interactions, there is no other mechanism in the
SM which can generate CP violating effects if $J_q=0$.  Note that in the
chiral limit, $m_u=m_d=m_s=0$, we do not generate CP violation even if
$J_q\ne0$.

In the literature, the lowest weak basis invariant is called Jarlskog
determinant \cite{Jarlskog:1985ht},
\begin{align}
  \begin{split}
    \det \left[H_u,H_d\right] = 2i & \left(m^2_t-m^2_c\right)
    \left(m^2_t-m^2_u\right) \left(m^2_c-m^2_u\right)
    \\
    & \left(m^2_b-m^2_s\right) \left(m^2_b-m^2_d\right)
    \left(m^2_s-m^2_d\right) J_q \ .
  \end{split}
\end{align}
which is equivalent to the Eq.~(\ref{eq:CPJ}).\footnote{For any
  $3\times3$ traceless Hermitian matrix $M$ one has: $\tr M^3 = 3
  \left|M\right|$.}  The Jarlskog determinant is only applicable to
the case of three generations, in contrast to the more general
invariant in Eq.~(\ref{CPINV}).

\bigskip

Now let us add a down quark isosinglet.  The gauge couplings to quarks
and their mass terms are ($i,j=1,2,3$ and $\alpha=1,2,3,4$):
\begin{subequations}
  \label{eq:lagrangian}
  \begin{align}
    \mathscr{L}_W^q & = -\frac{g}{\sqrt 2} \left( \bar u_{L i}
      \gamma^{\mu} d_{L i}\, W_\mu^{+} + \text{h.c.}  \right) - e
    J_\text{EM}^\mu A_\mu \nonumber
    \\
    & \quad -\frac{g}{2\cos\,\theta_{\mathrm W}} \left( \bar u_{L i}
      \gamma^\mu u_{L i}- \bar d_{L i} \gamma^\mu d_{L i} - 2
      \sin^2\theta_\text{W}\, J_\text{EM}^\mu \right) Z_{\mu} \,
    \label{eq:lagrangian-weak}
    \\[3pt]
    \mathscr{L}_M^q & = -\left( \bar u_{L i}\, M_u^{ij}\, u_{R j} +
      \bar d_{L i}\, M_d^{i\alpha}\, d_{R \alpha} + \bar d_{L 4}\,
      m_d^\alpha\, d_{R \alpha} \right) + \text{h.c.}
    \label{eq:lagrangian-mass}
  \end{align}
\end{subequations}
where the matrices $M_u$, $M_d$ and $m_d$ are of dimension $3 \times
3$, $3 \times 4$ and $1 \times 4$, respectively.  The electromagnetic
current is given by \mbox{$J_{\mathrm EM}^\mu = \frac{2}{3} \bar u
  \gamma^\mu u-\frac{1}{3} \bar d \gamma^\mu d$}.

The most general weak basis transformation consistent with the
Lagrangian of Eq.~(\ref{eq:lagrangian}) is:
\begin{align}
  \begin{pmatrix}
    u_{L i} \cr d_{L i}
  \end{pmatrix}
  & \longrightarrow U_L^{ij}
  \begin{pmatrix}
    u_{L j} \cr d_{L j}
  \end{pmatrix}
  , & u_{R i} &\longrightarrow \left(U^u_R\right)^{ij} u_{R j}\ , &
  d_{R \alpha} & \longrightarrow \left(U^d_R\right)^{\alpha\beta} d_{R
    \alpha}\ .
\end{align}
where $U_L$ and $U^u_R$ are $3\times3$ unitary matrices, while $U^d_R$
is $4\times4$.  Once we diagonalise the mass terms, the Lagrangian reads
\begin{align}
  \mathscr{L}_W & = - \frac{g}{\sqrt 2} \left[ \bar u_{L i}
    \gamma^{\mu} \left(V_\text{CKM}\right)_{i \alpha} d_{L \alpha}\,
    W_\mu^{+} + \text{h.c.}  \right] - e J_\text{EM}^\mu A_\mu
  \nonumber
  \\
  & \quad -\frac{g}{2\cos\,\theta_\text{W}} \left[ \bar u_{L i}
    \gamma^\mu u_{L i}- \bar d_{L \alpha} \gamma^\mu
    \left(V_\text{CKM}^\dagger V_\text{CKM}\right)_{\alpha\beta}\,
    d_{L \beta} - 2\sin^2\theta_\text{W}\, J_\text{EM}^\mu \right]
  Z_{\mu} \ ,\nonumber
  \\
  \mathscr{L}_M & = -\left( \bar u_{L i}\, D_{u i}\, u_{R i} + \bar
    d_{L \alpha}\, D_{d \alpha}\, d_{R \alpha} \right) + \text{h.c.} \
  ,
  \label{eq:lagrangian-new}
\end{align}
where $V_\text{CKM} = U^{u\,\dagger}_L U_L^{d}$ is a $3\times4$
matrix.  The number of independent phases which are related to CP
violation is, for $N=3$,
\begin{align}
  n_\text{CP} = N \left(N+1\right) - \frac{1}{2} N \left(N-1\right) -
  2N = \frac{1}{2} N (N-1) = 3 \ .
\end{align}

With the matrices as defined in Eq.~(\ref{eq:lagrangian-mass}) and
$H_u=M_u M_u^\dagger$, $H_d=M_d M_d^\dagger$, and $h_d=M_d
m_d^\dagger$, we can write down a set of weak basis invariants, 
\begin{align}
  I_1 & = \im \tr H_u H_d h_d h_d^\dagger \ , & %
  I_2 & = \im \tr H_u^2 H_d h_d h_d^\dagger \ , \nonumber
  \\[1mm]
  I_3 & = \im \tr H_u^2 \left[ H_u, H_d \right] h_d h_d^\dagger \ ,
  & %
  I_4 & = \im \tr H_u H_d^2 h_d h_d^\dagger \ , \nonumber
  \\[1mm]
  I_5 & = \im \tr H_u^2 H_d^2 h_d h_d^\dagger \ , & %
  I_6 & = \im \tr H_u^2 \left[ H_u, H_d^2 \right] h_d h_d^\dagger \ ,
  \nonumber
  \\[1mm]
  I_7 & = \im \tr H_u^2 H_d H_u H_d^2 \ ,
  \label{eq:invariants}
\end{align}
representing a set of necessary and sufficient conditions for having
CP invariance in the quark sector \cite{delAguila:1997vn}.

In our model, $H_d$ and $h_d$ read
\begin{align}
  H_d & =
  \begin{pmatrix}
    \abs{\mu_1}^2 + \tm_1^2 & \tm_1 \tm_2 & \tm_1 \tm_3 \cr \tm_1
    \tm_2 & \abs{\mu_2}^2 + \tm_2^2 & \tm_2 \tm_3 \cr \tm_1 \tm_3 &
    \tm_2 \tm_3 & \abs{\mu_3}^2 + \tm_3^2
  \end{pmatrix}
  , & h_d & =
  \begin{pmatrix}
    \mu_1 \tM_1 + \tm_1 \tM_4 \cr \mu_2 \tM_2 + \tm_2 \tM_4 \cr \mu_3
    \tM_3 + \tm_3 \tM_4
  \end{pmatrix}
  .
\end{align}
Since $H_u$ and $H_d$ are real, $I_7$ vanishes.  The remaining
invariants are in general different from zero; the dominant terms are
\begin{align}
  I_1 & = - m_t^2 \left(\tm_1^2+\tm_2^2\right) \tm_3 \tM_4 \im{\mu_3}
  \ , & I_2 & = m_t^2\, I_1 \ , \nonumber
  \\[1mm]
  I_3 & = - m_t^6 \left(\tm_1^2+\tm_2^2\right) \tm_3 \tM_3 \tM_4
  \im{\mu_3} \ , \nonumber
  \\[1mm]
  I_4 & = -m_t^2\left(\tm_1^2+\tm_2^2\right)
  \left(\tm_1^2+\tm_2^2+\tm_3^2+\mu_3^2\right) \tm_3 \tM_3 \tM_4
  \im{\mu_3} \ , & I_5 & = m_t^2\, I_4 \ , \nonumber
  \\[1mm]
  I_6 & = -m_t^6\left(\tm_1^2+\tm_2^2\right)
  \left(\tm_1^2+\tm_2^2+\tm_3^2+\mu_3^2\right) \tm_3 \tM_3 \tM_4
  \im{\mu_3} \ .
  \label{eq:quark-invariants}
\end{align}
Hence, CP is generally violated even by the presence of a single complex
parameter $\mu_3$.  Note that this case is not equivalent to the chiral
limit because both the charm and strange masses are different from zero,
$m_c\propto\mu_2$ and $m_s\sim\tm_2$ (albeit $\mu_2\ll\tm_2$).  As we
might expect, the invariants vanish if all quarks of the first and
second generation are massless.

Now we single out the heavy eigenstate with the rotations $V_4$, $U_4$.
While the action of $V_4$ leaves the invariants unaffected, $U_4$
strongly modifies them and reshuffles terms from one to the other.  In
fact after this transformation, $h_d$ vanishes to lowest order and
survives only at order $\mathcal{O}(v_{EW}^2/\tM^2)$; then in the new
basis all the invariants involving $h_d$, i.e., $I_1-I_6$ are suppressed
by $v_\text{EW}^2/\tM^2$ and vanish for $\tM\to\infty$.  On the other
hand $I_7$ is now non-vanishing and given by
\begin{align} \label{eq:eff-inv}
  I_7^\prime & = \im \tr H_u^2 H_d^\text{eff} H_u
  \left(H_d^\text{eff}\right)^2 \; ,
\end{align}
where $H_d^\text{eff}=\widehat{m} \widehat{m}^\dagger$ (see
Eq.~(\ref{eq:mhat})).  Note that $U_4$ also changes the weak
interactions,
\begin{align}
  \delta\mathscr{L}_W & = - \frac{g}{\sqrt{2}}\, \bar u_i \gamma^\mu
  \left(U_4-\mathbbm{1}\right)_{i4} d_4\, W_\mu^+ + \bar d_i \gamma^\mu
  \left( U_4^\dagger U_4 - \mathbbm{1} \right) _{i4} d_4\, Z_\mu +
  \text{h.c.}  ,
\end{align}
so we expect both CP violation and CKM unitarity violation from these
terms as well. However, the mass of the heavy state is
$\mathcal{O}\left(M_\text{GUT}\right)$ so that the contributions to
low-energy processes are suppressed by a factor
$M_\text{EW}/M_\text{GUT}$ and are negligible.

Hence, at the electroweak scale, we are left to consider the single
invariant
\begin{align}
  I_7^\prime & = \im \tr H_u^2 H_d^\text{eff} H_u H_d^\text{eff\,2} \; ,
\end{align}
which corresponds to the usual Jarlskog invariant $J_q$ for three
generations, but computed for the effective quark mass $\widehat{m}$.

\paragraph{Lepton Sector.}
As discussed above, we can ignore the heavy states for low-energy CP
violation and use the effective $3\times 3$ Yukawa matrices instead.

In the SM, extended by right-handed neutrinos, we have three mass
terms for the leptons,
\begin{align}
  \mathscr{L}_M^\ell & = -\left( \bar e_{Li}\, m_e^{ij}\, e_{Rj} +
    \bar \nu_{Li}\, m_D^{ij}\, \nu_{Rj} + \frac{1}{2}\, \nu_{Ri}^\top
    C\, m_N^{ij}\, \nu_{Rj}\right) + \text{h.c.}
\end{align}
In analogy to the quark sector, invariance of the mass terms under CP
transformation requires
\begin{align}
  U^\dagger m_e V & = m_e^\ast \;, & U^\dagger m_D W & = m_D^\ast
  \;, & W^\top m_N W & = -M_R^\ast \;,
\end{align}
where $U$, $V$, and $W$ are unitary matrices acting in flavour space.
Defining $h = m_D^\dagger m_D$ and $H = m_N^\dagger m_N$,
we obtain
\begin{align}
  W^\dagger h\, W & = h^\ast \ , & W^\dagger H\, W & = H^\ast \ .
\end{align}
Now we can write down the weak basis invariants
\begin{align}
  I_1^\ell & = \im \tr h\, H\, m_N^\ast h^\ast m_N , & I_2^\ell & =
  \im \tr h\, H^2 m_N^\ast h^\ast m_N , \nonumber
  \\
  I_3^\ell & = \im \tr h\, H^2 m_N^\ast h^\ast m_N H ;
  \label{eq:lepton-invariants}
\end{align}
for the three further invariants, substitute $\ovl{h} =
m_D^\dagger m_e\, m_e^\dagger\, m_D$ for $h$
\cite{Branco:2001pq}.
In the basis where the right-handed neutrino mass is diagonal, one
obtains
{\allowdisplaybreaks
  \begin{align}
    I_1^\ell & = M_1 M_2 \left(M_2^2-M_1^2\right) \im{h_{12}^2}
    \nonumber
    \\
    & \qquad + M_1 M_3 \left(M_3^2-M_1^2\right) \im{h_{13}^2} + M_2
    M_3 \left(M_3^2-M_2^2\right) \im{h_{23}^2} \;, \nonumber
    \\
    I_2^\ell & = M_1 M_2 \left(M_2^4-M_1^4\right) \im{h_{12}^2}
    \nonumber
    \\
    & \qquad + M_1 M_3 \left(M_3^4-M_1^4\right) \im{h_{13}^2} + M_2
    M_3 \left(M_3^4-M_2^4\right) \im{h_{23}^2} \;, \nonumber
    \\
    I_3^\ell & = M_1^3 M_2^3 \left(M_2^2-M_1^2\right) \im{h_{12}^2}
    \nonumber
    \\
    & \qquad + M_1^3 M_3^3 \left(M_3^2-M_1^2\right) \im{h_{13}^2} +
    M_2^3 M_3^3 \left(M_3^2-M_2^2\right) \im{h_{23}^2} \;. \nonumber
  \end{align}
}
If none of the $M_i$ vanish and there is no degeneracy, the vanishing of
$I_1$, $I_2$, and $I_3$ implies the vanishing of $\im{h_{12}^2}$,
$\im{h_{13}^2}$, and $\im{h_{23}^2}$ for CP invariance.

Note that in our model, $m_D$ stands for the effective $3\times 3$ part
of the Dirac neutrino mass matrix, $\ovl{m}_D$, as given in
Eq.~(\ref{eq:mDbar}).  Then we obtain from
Eq.~(\ref{eq:neutrino-dirac-triagonal}),
\begin{align}
  h_{12} & = A^\ast D \rb_1^2 + B^\ast E \rb_2^2 + C^\ast F \rb_3^2
  \;, \nonumber
  \\
  h_{13} & = A \rb_1^2 + B \rb_2^2 + C \rb_3^2 \;, 
  \nonumber\\
  h_{23} & = D^\ast \rb_1^2 + E^\ast \rb_2^2 + F^\ast \rb_3^2 \;. 
\end{align}
The coefficients $A,\ldots,F$ are displayed in
Eqs.~(\ref{eq:rhobar-a-f}).  They are generically complex, so we do not
expect CP to be conserved.

As in the quark sector, these invariants are rather general and give the
necessary conditions for the presence of CP violation.  On the other
hand, only few of the phases remain important also in the low-energy
limit.  In our case, to study the low-energy Dirac invariant, we can use
the analogue of the Jarlskog invariant,
\begin{align}
  J_\ell & = -\frac{1}{\mathscr{M}_\nu^2 \mathscr{M}_e^2}\, \tr
  \left[
    h^\nu_\text{eff}, h^e \right]^3 ,
\end{align}
as discussed in Section~\ref{se:oscillations}.  Here, $h^\nu_\text{eff}
= \left(m^\nu_\text{eff}\right)^\dagger m^\nu_\text{eff}$ and $\Delta
\mathscr{M}_\nu^2$ and $\Delta \mathscr{M}_e^2$ are the products of the
mass squared differences of the light neutrinos and charged leptons,
respectively.

\end{appendix}


\end{document}